\documentclass[aps,prb,amsmath,amssymb,reprint,superscriptaddress,showpacs,intlimits]{revtex4-1}
\usepackage{bm,latexsym,mathrsfs,enumerate,xcolor}
\usepackage[mathcal]{euscript}
\usepackage[breaklinks=true,unicode=true,urlcolor = blue,colorlinks = true,citecolor = blue,linkcolor = blue]{hyperref}
\usepackage{pgf,tikz}
\usepackage{graphicx}
\usepackage{tabularx}
\newcolumntype{C}[1]{>{\centering\arraybackslash}p{#1}}
\renewcommand{\vec}[1]{\bm{#1}}
\begin{document}
\preprint{\textcolor[rgb]{0.00,0.50,0.75}{{\texttt{Draft v2.32}}}}
\title{Torsion induced effects in magnetic nanowires}

\author{Denis D. Sheka}
\email[Corresponding author:]{sheka@univ.net.ua}
\affiliation{Taras Shevchenko National University of Kyiv, 01601 Kyiv, Ukraine}

\author{Volodymyr P. Kravchuk}
\email{vkravchuk@bitp.kiev.ua}
\affiliation{Bogolyubov Institute for Theoretical Physics, 03143 Kyiv, Ukraine}

\author{Kostiantyn V. Yershov}
\email{yershov@bitp.kiev.ua}
\affiliation{Bogolyubov Institute for Theoretical Physics, 03143 Kyiv, Ukraine}
\affiliation{National University of ``Kyiv-Mohyla Academy", 04655 Kyiv, Ukraine}

\author{Yuri Gaididei}
\email{ybg@bitp.kiev.ua}
\affiliation{Bogolyubov Institute for Theoretical Physics, 03143 Kyiv, Ukraine}

\date{June 13, 2015}

%
%%%%%%%%%%%%%%%%%%%%%%%%%%%%%%%%%%%%%%%%%%%%%%%%%%%%%%%%%%%%%%%%%%%%%70
%
%         ABSTRACT
%
%%%%%%%%%%%%%%%%%%%%%%%%%%%%%%%%%%%%%%%%%%%%%%%%%%%%%%%%%%%%%%%%%%%%%70
%
\begin{abstract}
Magnetic helix wire is one of the most simple magnetic systems which manifest properties of both curvature and torsion. There exist two equilibrium states in the helix wire with easy-tangential anisotropy: a quasi-tangential magnetization distribution in case of relatively small curvatures and torsions, and an onion state in opposite case. In the last case the magnetization is close to tangential one, deviations are caused by the torsion and curvature. Possible equilibrium magnetization states in the helix magnet with different anisotropy directions are studied theoretically. The torsion also essentially influences the spin-wave dynamics, acting as an effective magnetic field. Originated from the curvature induced effective Dzyaloshinskii interaction, this magnetic field leads to the coupling between the helix chirality and the magnetochirality, it breaks mirror symmetry in spin-wave spectrum. All analytical predictions on magnetization statics an dynamics are well confirmed by the direct spin lattice simulations.
\end{abstract}
%\keywords{classical spin models,curvature,nanowire, nanoshell}
%\submitto{\jpa}
% 75.30.Et	Exchange and superexchange interactions (see also 71.70.Gm Exchange interactions)
% 75.75.-c	Magnetic properties of nanostructures
% 75.78.-n	Magnetization dynamics

\maketitle
%
%
%%%%%%%%%%%%%%%%%%%%%%%%%%%%%%%%%%%%%%%%%%%%%%%%%%%%%%%%%%%%%%%%%%%%%70
%
%
\section{Introduction}
\label{sec:intro}

During the past few years there is a growing interest to curvature effects in physics of nanomagnetism. A crucial aspect of the interest is caused by recent achievements in nanotechnologies of flexible, stretchable and printable magnetoelectronics (see Ref.~\onlinecite{Makarov13a} and references therein). Effects of the curvature on the magnetization structure in nanomagnetic particles of nontrivial geometry were studied for cylinders, \cite{Saxena98,Carvalho-Santos13a} torus, \cite{Carvalho-Santos08}, half-spheres\cite{Yoo15}, spherical shells, \cite{Kravchuk12a} hemispherical caps, \cite{Streubel12,Sheka13b} cylindrical capped nanomembranes, \cite{Streubel12b}, cone shells \cite{Gaididei14,Sheka15}, and paraboloidal shells \cite{Carvalho-Santos15}. Chiral and curvature effects with account of the nonlocal dipolar interaction were discussed for cylinder nanotubes. \cite{Landeros10,Otalora12a,Yan12}

Very recently we have developed fully three dimensional (3D) approach for studying statics and dynamics of thin magnetic shells and wires of arbitrary shape. \cite{Gaididei14,Sheka15} This approach gives a possibility to derive the energy for arbitrary curves and surfaces and arbitrary magnetization vector fields on the assumption that the anisotropy contribution greatly exceeds the dipolar and other weak interactions, i.e. for hard magnets. We have shown \cite{Sheka15} that due to the curvature two additional effective magnetic interactions originate from the exchange term: (i) curvature induced effective anisotropy, which is bilinear with respect to curvature and torsion and (ii) curvature induced effective Dzyaloshinskii interaction, which is linear with respect to curvature and torsion. This novel approach open doors for studying several perspective directions in nanomagnets, including topologically induced patterns\cite{Kravchuk12a,Pylypovskyi14d} and magnetochiral effects\cite{Pylypovskyi14d,Sheka15}.

The simplest system which displays both the properties of the curvature and torsion is a helix wire, which is characterized by constant curvature and torsion. The interest to such a geometry is motivated by recent experiments on rolled--up ferromagnetic microhelix coils. \cite{Smith11,Smith11a} Depending on the anisotropy direction different artificial complex helimagnetic--like configurations were experimentally realized: hollow--bar--, \mbox{corkscrew--,} and radial--magnetized 3D micro--helix coils. \cite{Smith11} Rolled magnetic structures are now widely discussed in the context of possible application in flexible and stretchable magnetoelectronic devices, \cite{Streubel14} in particular, rolled-up GMR sensors, \cite{Moench11} for magnetofluidic applications,  spin-wave filters, \cite{Balhorn10,Balhorn13} and microrobots \cite{Solovev10}. Helix coil magnetic structures  have the potential to be used a variety of bioapplication areas, such as in medical procedures, cell biology, or lab--on--a--chip. \cite{Peyer12}

In the current study we apply our theory \cite{Sheka15} aimed to describe magnetization statics and linear dynamics in the helix wire. We analyze equilibrium states for different types of magnetocrystalline anisotropy. The equilibrium state is determined by the relationship between the curvature, torsion and the anisotropy strength: we describe possible magnetization distributions analytically. For three types of anisotropy (easy-tangential, easy-normal and easy-binormal) we compute phase diagrams of possible equilibrium states. In each of these cases the equilibrium state is either onion one (high curvatures) or anisotropy-aligned state (for small ones), these results are summarized in Fig.~\ref{fig:helix_phd}. For example, in the most interesting case of easy-tangential anisotropy a quasi-tangential magnetization distribution appears for strong enough anisotropy, see Fig.~\ref{fig:helix}(a,b). We show that pure tangential magnetization distribution is impossible. The deviation from the tangential state is determined by the the curvature and torsion; besides there exists the coupling between the helix chirality and the magnetochirality of magnetization distribution. 

We study the problem of spin wave dynamics in the helix wire. Our analysis shows that the curvature and torsion act on magnons in two ways: besides the standard potential scattering of magnons, there appears an effective torsion induced magnetic field. The vector potential of effective field is mainly determined by the product of the torsion and the magnetochirality. The origin of this field is the curvature induced effective Dzyaloshinskii interaction. \cite{Sheka15} Finally, the torsion breaks the symmetry of spin wave spectrum with respect to the direction of spin wave propagation, see Fig.~\ref{fig:helix_dispersion}. This effect is completely analogous to the effect of asymmetry of magnon dispersion due to the natural Dzyaloshinskii interaction in magnetic films.\cite{Zakeri10,Cortes-Ortuno13,Zakeri14}

The paper is organized as follows. In Sec.~\ref{sec:model} we introduce the model of the curved wire and discuss different anisotropy-aligned states. The model of the helix wire appears in Sec.~\ref{sec:equilibrium}. Equilibrium magnetization distributions are describe analytically for the easy--tangential helix wire: the quasi-tangential state (see Sec.~\ref{sec:helix}) and the onion one (see Sec.~\ref{sec:onion}). The phase diagram of energetically preferable states appears in Sec.~\ref{sec:phase-diagram}. The problem of spin-wave dynamics is discussed in Sec.~\ref{sec:sw}. In Sec.~\ref{sec:anis} we study statics and linear dynamics for helix wires with other anisotropy orientations: the easy-normal anisotropy (Sec.~\ref{sec:EN}) and easy-binormal one (Sec.~\ref{sec:EB}). We verify our theory by numerical simulations of the helix-shaped chain of discrete magnetic moments in Sec.~\ref{sec:simulations}. In Section \ref{sec:discussion} we present final remarks and discuss possible perspectives and generalizations, in particular, how to take into account magnetostatics effects. Some details about the computation of the onion state are presented in Appendix~\ref{app:onion}.

\section{The model of a curved wire}
\label{sec:model}

We consider a model of a curved cylindrical wire. Let $\vec{\gamma }(s)$ be a 1D curve embedded in 3D space $\mathbb{R}^3$ with $s$ being the arc length coordinate. It is convenient to use Frenet--Serret reference frame with basic vectors $\vec{e}_\alpha $:
\begin{equation} \label{eq:FNT-basis}
\vec{e}_{\text{\sc{t}}} = \vec{\gamma }', \qquad \vec{e}_{\text{\sc{n}}} = \frac{\vec{e}_{\text{\sc{t}}}'}{\left|\vec{e}_{\text{\sc{t}}}' \right|}, \qquad \vec{e}_{\text{\sc{b}}} = \vec{e}_{\text{\sc{t}}}\times \vec{e}_{\text{\sc{n}}}
\end{equation}
with $\vec{e}_{\text{\sc{t}}}$ being the tangent, $\vec{e}_{\text{\sc{n}}}$ being the normal, and $\vec{e}_{\text{\sc{b}}}$ being binormal to $\vec{\gamma }$. Here and below the prime denotes the derivative with respect to the arc length $s$ and Greek indices $\alpha ,\beta$ numerate curvilinear coordinates (TNB--coordinate) and curvilinear components of vector fields. The relation between $\vec e_\alpha'$ and $\vec{e}_\alpha $ is determined by Frenet--Serret formulas:
\begin{equation} \label{eq:Frenet-Serret}
\vec{e}_\alpha ' = F_{\alpha \beta }\vec{e}_\beta , \qquad \left\|F_{\alpha \beta } \right\| =
\begin{Vmatrix}
0 & \kappa & 0 \\
-\kappa & 0 & \tau  \\
0 &- \tau  & 0
\end{Vmatrix}
.
\end{equation}
Here $\kappa $ and $\tau $ are the curvature and torsion of the wire, respectively.

The wire of a finite thickness $h$ can be defined as the following space domain
\begin{equation} \label{eq:3D-curve}
\vec{r}(s,u,v) = \vec{\gamma }(s) + u \vec{e}_{\text{\sc{n}}} + v \vec{e}_{\text{\sc{b}}},
\end{equation}
where $u$ and $v$ are coordinates within the wire cross section ($|u|, |v|\lesssim h$).

Let us describe the magnetic properties of the wire.  The magnetic energy of the wire can collect different contributions such as energies of exchange interaction, anisotropy, and dipolar one. We start our analysis with the case of a \emph{hard} magnet where the anisotropy contribution greatly exceeds the dipolar and other weak interactions. For such hard magnets a quality factor\cite{Hubert98}
\begin{equation} \label{eq:Q-factor}
Q\equiv \frac{K}{2\pi M_s^2}
\end{equation}
is supposed to be large; here $K>0$ is the constant of magnetocrystalline anisotropy and $M_s$ is the saturation magnetization.

We assume the magnetization spatial one-dimensionality, which can be formalized as $\vec m=\vec m(s,t)$. This assumption is appropriate for the cases when the thickness $h$ does not exceed the characteristic magnetic length $w = \sqrt{\mathcal{A}/K}$ with $\mathcal{A}$ being the exchange constant. The wire thickness is also supposed to be small in comparison with radii of curvature and torsion. Therefore our model provides an adequate picture under the following assumptions:
\begin{equation} \label{eq:hard-validity}
h \lesssim w \ll \frac{1}{\kappa},\ \frac{1}{\tau}, \qquad Q\gg1.
\end{equation}

That is why in the current study one can restrict ourself to the consideration of Heisenberg magnets with the energy
\begin{equation} \label{eq:enegry}
\begin{split}
E &= \mathcal{A} S \int \mathrm{d}s \bigl(\mathscr{E}_{\text{ex}} + \mathscr{E}_{\text{an}} \bigr),\\
\mathscr{E}_{\text{ex}} &= -\vec{m}\cdot\vec{\nabla}^2\vec{m},\qquad \mathscr{E}_{\text{an}}  = - \frac{\left(\vec{m}\cdot\vec{e}_{\text{an}}\right)^2}{w^2}
\end{split}
\end{equation}
with $\vec{e}_{\text{an}}$ being the unit vector along the anisotropy axis, and $S$ being the cross-section area.

Typically, orientation of the anisotropy axis $\vec{e}_{\text{an}}$ is determined by the wire geometry, e.g. it can be tangential to the wire, \cite{Smith11} which means in general complicated spatial dependence due to the curvilinear geometry. Therefore it is convenient to represent the energy of the magnet in the curvilinear reference frame \eqref{eq:FNT-basis}, where $\mathscr{E}_{\text{an}}$ has a simplest form. For an arbitrary thin wire the exchange energy density can be presented as follows \cite{Sheka15}
\begin{equation} \label{eq:energy-Frenet}
\begin{split}
\mathscr{E}_{\text{ex}} &= \mathscr{E}_{\mathrm{ex}}^{0} + \mathscr{E}_{\mathrm{ex}}^{A} + \mathscr{E}_{\mathrm{ex}}^{D}, \quad \mathscr{E}_{\mathrm{ex}}^{0} = \left| \vec{m}'\right|^2,\\
\mathscr{E}_{\mathrm{ex}}^{A} &= K_{\alpha \beta }m_\alpha m_\beta ,\quad \mathscr{E}_{\mathrm{ex}}^{D} = F_{\alpha \beta } \left(m_\alpha  m_\beta ' - m_\alpha' m_\beta  \right).
\end{split}
\end{equation}
Here the first term $\mathscr{E}_{\mathrm{ex}}^{0}$ describes the common isotropic part of exchange expression which has the same form as for the straight wire. The second term $\mathscr{E}_{\mathrm{ex}}^{A}$ describes an effective anisotropy interaction, where the components of the tensor $K_{\alpha \beta } = F_{\alpha \nu }F_{\beta \nu }$ are bilinear with respect to the curvature $\kappa $ and the torsion $\tau $. This term is similar to the ``geometrical potential''. \cite{Costa81} Note that a curvature caused ``geometric'' effective magnetic field was considered recently for curved magnonic waveguides. \cite{Tkachenko12} The last term $\mathscr{E}_{\mathrm{ex}}^{D}$ in the exchange energy functional is the curvature induced effective Dzyaloshinskii interaction, which is linear with respect to curvature and torsion. We will see below that this effective interaction causes an effective magnetic field; namely this interaction is responsible for the magnetochiral effects.

%==================================================================\
\begin{table}
\begin{tabular}{p{0.27\columnwidth} C{0.2\columnwidth} p{0.3\columnwidth} p{0.2\columnwidth} }\hline\hline
Anisotropy   & Anisotropy   & \multicolumn{2}{c}{Magnetization states}\\
type         & axis         & \multicolumn{2}{c}{in a helix wire}\\
& $\vec{e}_{\text{an}}$ & Equilibrium states & Orientation according to Ref.~\onlinecite{Smith11} \\\hline
Easy--tangential & $\vec{e}_{\text{\sc{t}}}$  & quasi-tangential and onion  & corkscrew \\
Easy--normal     & $\vec{e}_{\text{\sc{n}}}$  & normal and onion & radial \\
Easy--binormal   & $\vec{e}_{\text{\sc{b}}}$  & quasi-binormal and onion    & hollow--bar \\ \hline\hline
\end{tabular}
\caption{Types of equilibrium magnetization states for various uniaxial anisotropies in a helix-shaped magnetic wire.}
\label{tab:anisotropy}
\end{table}
%==================================================================/

We consider three types of curvilinear uniaxial anisotropy which correspond to possible curvilinear directions \eqref{eq:FNT-basis}, see Table~\ref{tab:anisotropy}: (i) an easy--tangential anisotropy corresponds to the anisotropy axis $\vec{e}_{\text{an}}$ directed along $\vec{e}_{\text{\sc{t}}}$, where the anisotropy interaction tries to orient the magnetization along the curve. Note that in soft magnets such kind of anisotropy appears effectively as a shape anisotropy caused by the dipolar interaction.\cite{Slastikov12} (ii) An easy--normal anisotropy is determined by the normal vector $\vec{e}_{\text{\sc{n}}}$. (iii) An easy--binormal anisotropy direction corresponds to the binormal basic vector $\vec{e}_{\text{\sc{b}}}$.

All three types of anisotropic magnets can be realized experimentally: In straight nanostrips/nanowires the anisotropy can have well--defined uniaxial directions, e.g., in-plane along the strip,  in-plane perpendicularly to the strip, or out-of-plane, which corresponds to the uniformly magnetized samples in the corresponding direction. Using the coiling process,\cite{Smith11} it is possible to obtain 3D microhelix coil strips with different magnetization orientation: corkscrew-, radial-, and hollow-bar-magnetized, see Table~\ref{tab:anisotropy} to get a link between  anisotropy type and the magnetization orientation.

For the further analysis it is convenient to introduce the angular parametrization of the magnetization unit vector $\vec{m}$ using the local Frenet--Serret reference frame:
\begin{equation} \label{eq:m-parametrization}
\vec{m} = \sin\theta \cos\phi \, \vec{e}_{\text{\sc{t}}} + \sin\theta \sin\phi \,\vec{e}_{\text{\sc{n}}} + \cos \theta \, \vec{e}_{\text{\sc{b}}},
\end{equation}
where angular variables $\theta $ and $\phi $ depend on both spatial and temporal coordinates. Then the energy density \eqref{eq:energy-Frenet} reads: \cite{Sheka15}
\begin{subequations} \label{eq:energy-angular}
%\begin{equation} \label{eq:Eex-angular}
\begin{align} \label{eq:Eex-angular}
&\mathscr{E}_{\text{ex}} \!= \left[\theta '\!-\tau \sin\phi \right]^2\! + \left[\sin\theta  (\phi '+\kappa )-  \tau \cos\theta \cos\phi  \right]^2\!\!\!,\\
%\end{equation}
\label{eq:Ean-angular} %
&\mathscr{E}_{\text{an}}^{\mathrm{ET}} = -\frac{\sin^2\theta \cos^2\phi }{w^2},\quad
\mathscr{E}_{\text{an}}^{\mathrm{EN}} = -\frac{\sin^2\theta \sin^2\phi }{w^2},\nonumber\\
&\mathscr{E}_{\text{an}}^{\mathrm{EB}} = -\frac{\cos^2\theta }{w^2}.
\end{align}
\end{subequations}
Here $\mathscr{E}_{\text{an}}^{\mathrm{ET}}$, $\mathscr{E}_{\text{an}}^{\mathrm{EN}}$, and $\mathscr{E}_{\text{an}}^{\mathrm{EB}}$ denotes anisotropy energy densities of easy--tangential, easy--normal, and easy--binormal types, respectively.

The magnetization dynamics follows the Landau--Lifshitz equation. In terms of angular variables $\theta $ and $\phi $ these equations read
\begin{equation} \label{eq:LL}
\frac{M_s}{\gamma_0} \sin\theta  \partial_t \phi  = \frac{\delta  E}{\delta  \theta }, \quad -\frac{M_s}{\gamma_0 } \sin\theta  \partial_t \theta  = \frac{\delta  E}{\delta  \phi }
\end{equation}
with $M_s$ being the saturation magnetization and $\gamma_0$ being the gyromagnetic ratio.

\section{Equilibrium magnetization states of a helix wire with easy--tangential anisotropy}
\label{sec:equilibrium}

%==================================================================\
\begin{figure}
\begin{center}
\includegraphics[width=\columnwidth]{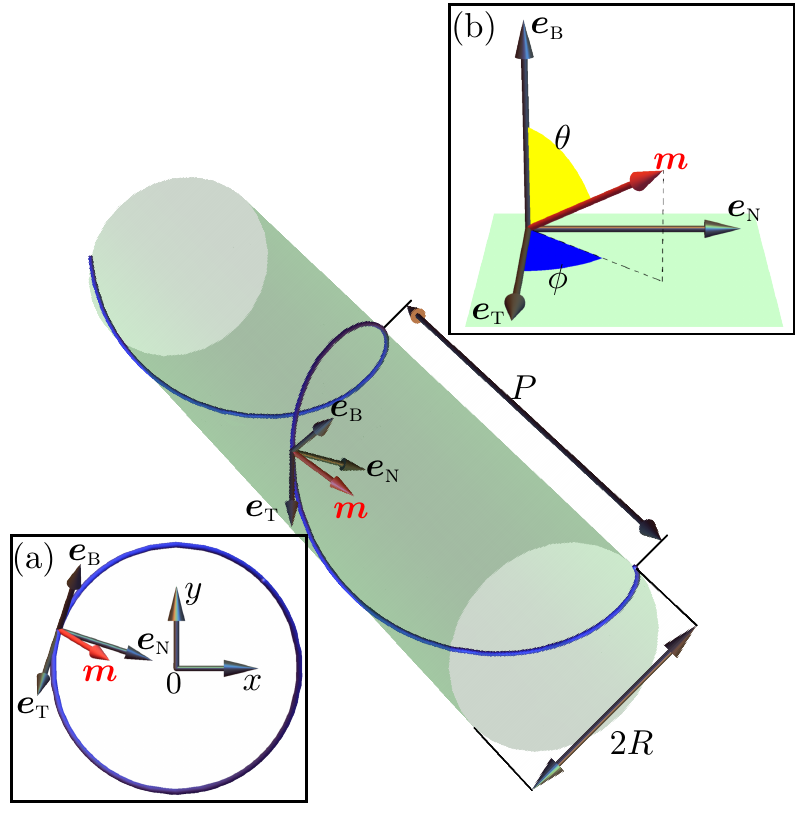}
\end{center}
\caption{(Color online)
Schematics of the helix wire of the radius $R$ and the pitch $P$. (a) Arrangement of the curvilinear Frenet--Serret reference frame $(\vec{e}_{\text{\sc{t}}}, \vec{e}_{\text{\sc{n}}}, \vec{e}_{\text{\sc{b}}})$ from the front view. (b) Arrangement of the magnetization angles $\theta$ and $\phi$ with respect to the magnetization unit vector $\vec{m}$.
}
\label{fig:helix_schematics}
\end{figure}
%==================================================================/

We study the curvilinear effects using the helix geometry, which is the simplest geometry which manifests the properties of both curvature and torsion. A typical parameterization of the helix wire reads
\begin{subequations} \label{eq:helix-param}
\begin{equation} \label{eq:helix-param-Rp}
\vec{\gamma }(\chi) = \hat{\vec x} R\cos \chi + \hat{\vec y} R\sin \chi +\hat{\vec z} p\chi,
\end{equation}
where $R$ is the helix radius, $p=P/(2\pi )$ with $P$ being the pitch of the helix, and $\chi$ is azimuthal angle of a cylindrical frame of reference with $\hat{\vec z}$-axis aligned along the helix axis, see Fig.~\ref{fig:helix_schematics}. The helix has the constant curvature $\kappa=R/(R^2+p^2)$ and the torsion $\tau=p/(R^2+p^2)$. For the further analysis it is instructive to rewrite \eqref{eq:helix-param-Rp} as a function of the arc length $s$ and in terms of curvature and torsion
\begin{equation} \label{eq:helix-param-kappatau}
\begin{split}
\vec{\gamma }(s) &= \hat{\vec x} \kappa  s_0^2 \cos \left(\frac{s}{s_0}\right) + \hat{\vec y}  \kappa  s_0^2 \sin \left(\frac{s}{s_0}\right) + \hat{\vec z} s_0 \tau  s,\\  s_0 &= \frac{1}{\sqrt{\kappa ^2+\tau ^2}}.
\end{split}
\end{equation}
\end{subequations}
One has to notice a one-to-one correspondence between $(R,p)$--parametrization \eqref{eq:helix-param-Rp} and $(\kappa,\tau)$--one \eqref{eq:helix-param-kappatau}.

In order to derive the explicit form of Landau--Lifshitz equations, we substitute the energy functional \eqref{eq:energy-angular} into the Landau--Lifshitz equations \eqref{eq:LL}:
\begin{equation} \label{eq:LLE-helix}
\begin{split}
&- \frac{M_s}{2\gamma_0 \mathcal{A}}\sin \theta \partial_t \phi  = \tau \cos \phi  \left( \kappa  \cos2\theta  - 2 \partial_s \phi  \sin^2\theta \right)\\
&+ \partial_{ss} \theta -\sin \theta  \cos \theta  \left[\left(\kappa  + \partial_s \phi \right)^2\!\! -\tau ^2\! \cos^2\phi  \right]  - \frac12 \frac{\partial \mathscr{E}_{\text{an}}}{\partial \theta },\\
&\frac{M_s}{2\gamma_0\mathcal{A}}\sin \theta \partial_t \theta  = \sin \theta \cos \theta  \left[2 \partial_s \theta  \left( \kappa  + \partial_s \phi \right)- \kappa  \tau  \sin \phi \right]\\
&+ \sin^2\!\theta  \left[\partial_{ss}\phi  + 2\tau \partial_s \theta \cos \phi  - \tau ^2\!\sin \phi \cos \phi  \right] - \frac12 \frac{\partial \mathscr{E}_{\text{an}}}{\partial \phi },
\end{split}
\end{equation}
where $\mathscr{E}_{\text{an}}$ is the density of the anisotropy energy, see \eqref{eq:Ean-angular}.

We are mostly interested in the case of easy--tangential anisotropy, which is typical for the wires. In this case the anisotropy energy density has the form $\mathscr{E}_{\text{an}}^{\mathrm{ET}}$, see \eqref{eq:Ean-angular}.

First we discuss the  limit case $\tau =0$ (ring wire instead of the helix). For any plane curve the energy functional \eqref{eq:energy-angular} with easy--tangential or easy--normal anisotropy is minimized by the plane magnetization distribution, $\theta _0=\pi /2$. The energy minimization in respect to $\phi $ results in the pendulum equation
\begin{equation} \label{eq:pendulum}
\varkappa ^2 \partial_{\chi \chi }\phi  -\sin\phi \cos\phi =0, \qquad \varkappa =\kappa  w
\end{equation}
with $\varkappa$ being the reduced curvature.

The equilibrium magnetization state of a ring is a homogeneous (in the curvilinear reference frame) vortex state $\phi^{\mathrm{vor}}$ in case of relatively small reduced curvature $\varkappa <\varkappa _0\approx 0.657$ and inhomogeneous onion solution $\phi^{\mathrm{on}}$ for $\varkappa >\varkappa _0$ \cite{Sheka15}
\begin{equation} \label{eq:vortex-onion}
\phi ^{\mathrm{vor}}=0,\pi ,  \qquad \phi^{\mathrm{on}} = \frac{\pi }{2}-\mathrm{am}(x,k),\; x=\frac{2\chi}{\pi}\mathrm{K}(k).
\end{equation}
Here $\mathrm{am}(x,k)$ is the Jacobi amplitude \cite{NIST10} and the modulus $k$ is determined by condition
\begin{equation} \label{eq:modulus}
2\varkappa  k\mathrm{K}(k)=\pi
\end{equation}
with $\mathrm{K}(k)$ being the complete elliptic integral of the first kind. \cite{NIST10}

\subsection{Quasi-tangential state}
\label{sec:helix}

%==================================================================\
\begin{figure}
\begin{center}
\includegraphics[width=\columnwidth]{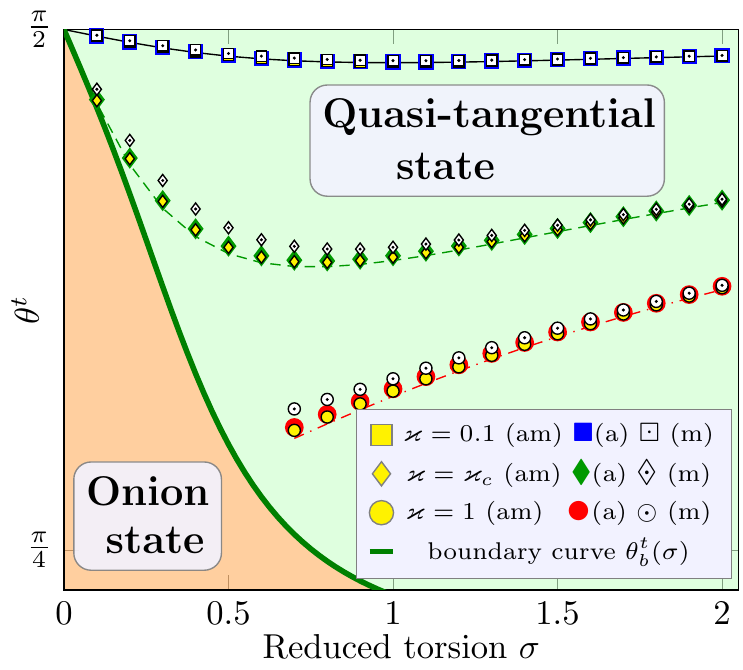}
\end{center}
\caption{(Color online) 
Equilibrium magnetization distribution in the quasi-tangential state of the helix wire with $\mathcal{C}=+1$. Lines correspond to the analytics, see Eq.~\eqref{eq:helix-gs-1}. Symbols correspond to simulations: (a) anisotropic Heisenberg magnets [see \eqref{eq:tot-energy-Heisenberg}] ($Q=2$, $w=\ell$), (am) wires with account of dipolar interaction [see \eqref{eq:tot-energy-eq-mod}, \eqref{eq:eff-kappa-sigma}] ($Q=2$, $w^{\text{eff}}=2\ell/\sqrt{5}$), and (ms) isotropic wires with account of dipolar interaction [see \eqref{eq:tot-energy-eq-mod},\eqref{eq:eff-kappa-sigma}] ($Q=2$, $w^{\text{eff}}=2\ell$). The boundary curve $\theta_b^t(\sigma)$ corresponds to \eqref{eq:theta-b}.	
}
\label{fig:helix_gs}
\end{figure}
%==================================================================/

Now we consider the helix wire with a finite torsion, $\tau\neq0$. Similar to the case of a ring wire, discussed above, we look for the homogeneous (in the curvilinear reference frame) solution. Such kind of solutions is possible due to the constant curvature $\kappa $ and the torsion $\tau $. We can easily solve the static equations, see Eq.~\eqref{eq:LLE-helix}, using the substitution $\theta (s)=\theta ^t$ and $\phi (s)=\phi ^t$:
\begin{equation} \label{eq:helix-gs}
\tan 2 \theta ^t = -\frac{2 \mathcal{C}\sigma \varkappa }{1-\varkappa ^2+\sigma ^2}, \qquad \phi ^t=0, \pi,
\end{equation}
where $\mathcal{C}=\cos\phi^t=\pm1$, the quantity $\sigma \equiv w\tau $ is the reduced torsion. Explicitly for magnetization angles we get
\begin{equation} \label{eq:helix-gs-1}
\begin{split}
& \theta ^t = \frac{\pi }{2} -  \arctan\frac{2\mathcal{C}\sigma \varkappa }{V_0}, \qquad \phi ^t=0, \pi,\\
& V_0 = 1+\sigma ^2-\varkappa ^2+V_1,\\
& V_1 = \sqrt{(1-\varkappa ^2+\sigma ^2)^2+4\varkappa ^2\sigma ^2}.
\end{split}
\end{equation}
The dependence $\theta ^t(\varkappa ,\sigma )$ is presented in Fig.~\ref{fig:helix_gs}.

%==================================================================\
\begin{figure}
	\begin{center}
		\includegraphics[width=\columnwidth]{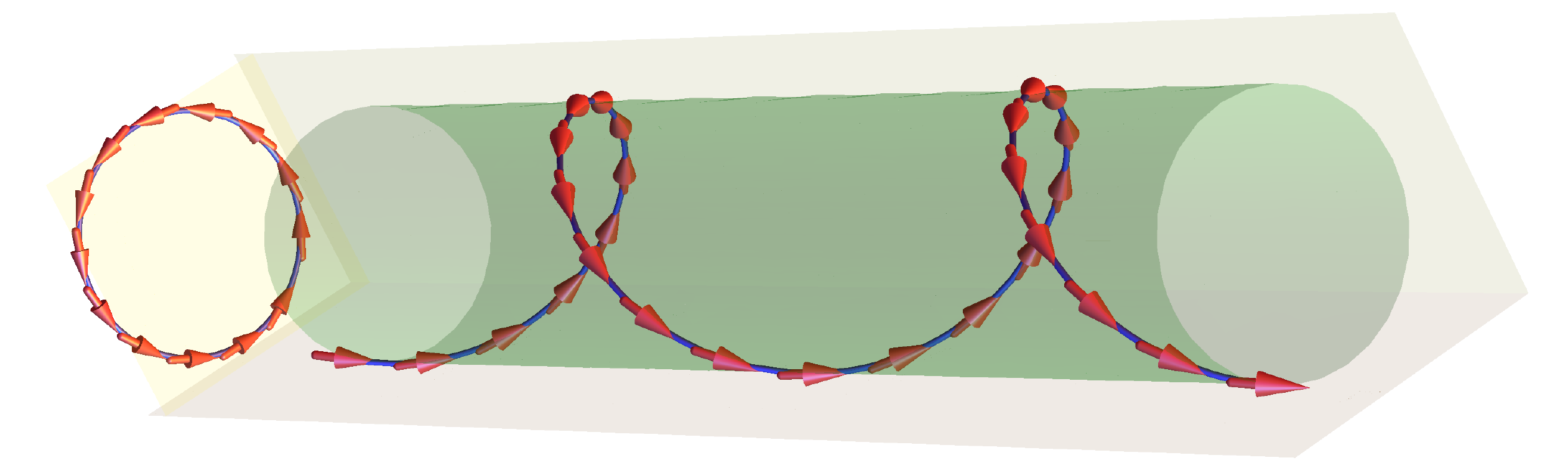}\\
		(a) Quasi-tangential state: $\varkappa =\sigma =0.1$\\
		\includegraphics[width=\columnwidth]{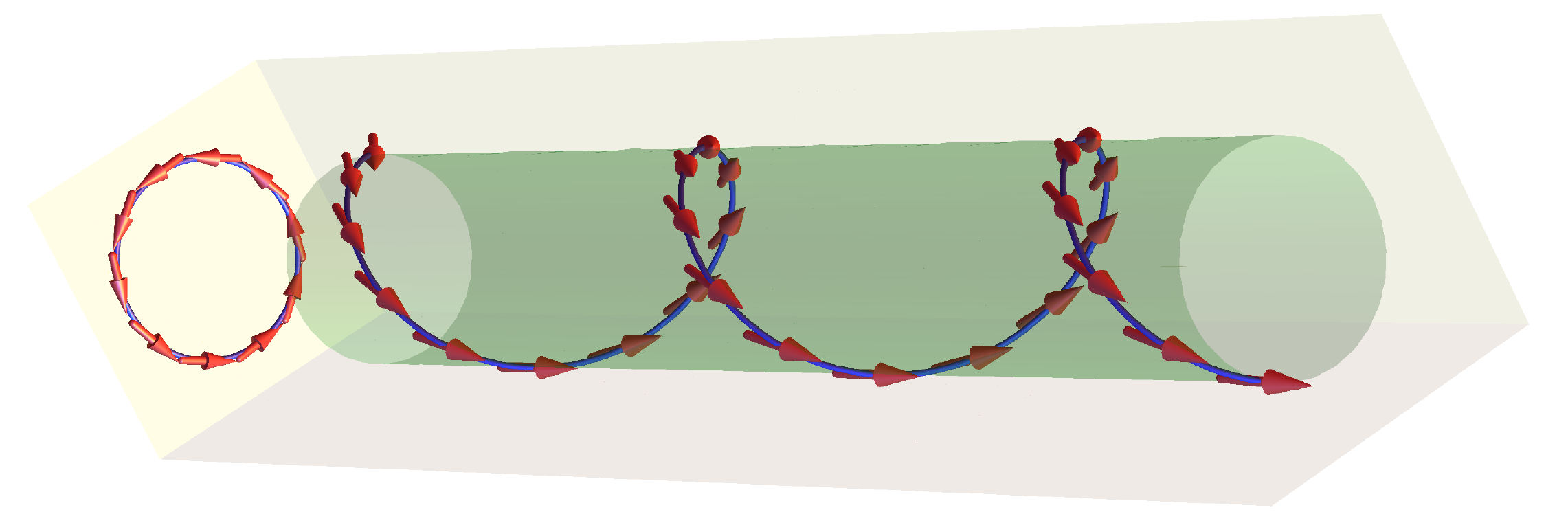}\\
		(b)  Quasi-tangential state: $\varkappa =\sigma =\varkappa_0\approx 0.657$\\
		\includegraphics[width=\columnwidth]{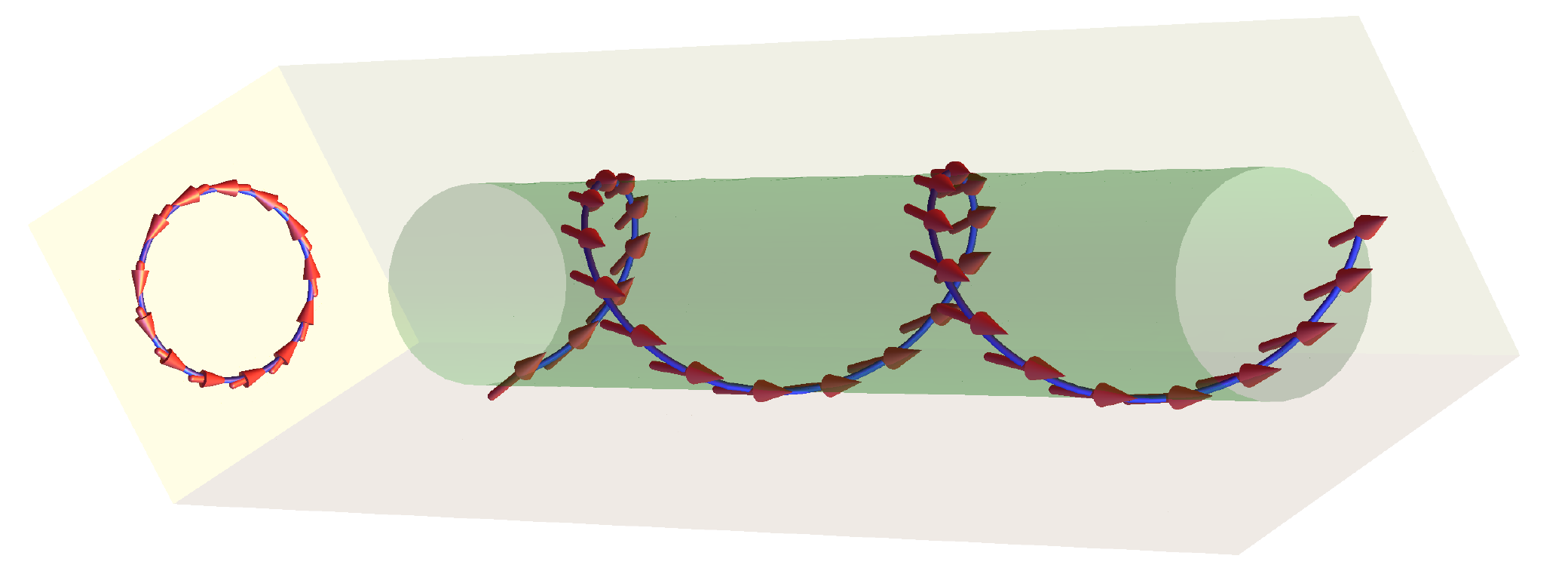}\\
		(c) Quasi-tangential state: $\varkappa =\sigma =1$\\
		\includegraphics[width=\columnwidth]{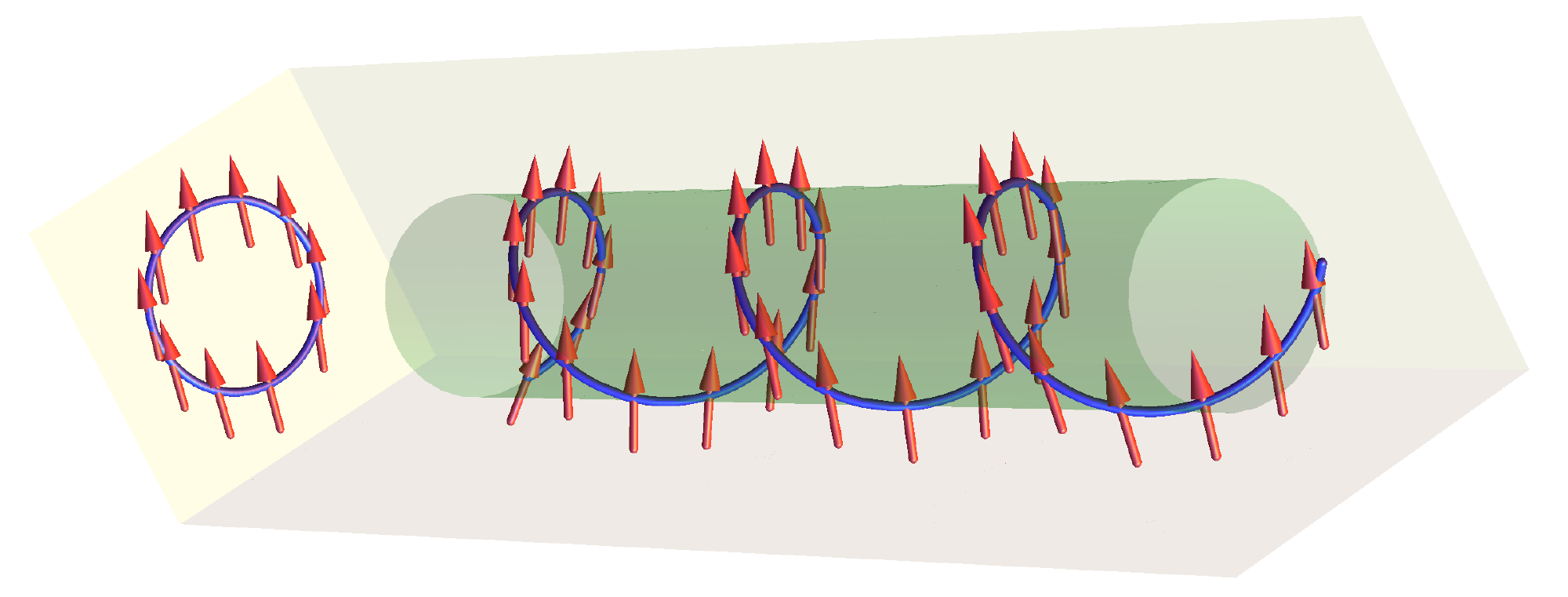}\\
		(d) Onion state: $\varkappa = 1.5$, $\sigma =1$\\
	\end{center}
	\caption{(Color online) Magnetization distributions in the helix wire with $\mathcal{C}=+1$ and easy-tangential anisotropy according to simulations data, see Sec.~\ref{sec:equilibrium-simulations}.
	}
	\label{fig:helix}
\end{figure}
%==================================================================/

In the limit case of very small curvature and torsion ($\varkappa ,\sigma \ll1$), the magnetization distribution becomes almost tangential, see Fig.~\ref{fig:helix}(a) with the asymptotic behavior
\begin{equation} \label{eq:helix-asymp}
\theta ^t \approx \frac{\pi }{2}- \mathcal{C}\sigma \varkappa , \qquad \text{for $\varkappa ,\sigma \ll1$}.
\end{equation}
That is why we refer to the state \eqref{eq:helix-gs-1} as to the \emph{quasi--tangential} state. Such a state is an analogue of the vortex state for the case of the torsion presence.

Even in the strong anisotropic case the magnetization deviates from the tangential distribution: the inclination angle depends on the sign of $\mathcal{C}\sigma $. One can interpret the sign of $\sigma $ as the helix chirality (different for right handed helix when $\sigma >0$ and left--handed one when $\sigma <0$); the quantity $\mathcal{C}$ can be interpreted as the magnetochirality, hence on can say about coupling between the two chiralities.

The energy density \eqref{eq:energy-angular} of the quasi--tangential state \eqref{eq:helix-gs-1} reads
\begin{equation} \label{eq:helix-gs-energy}
\mathscr{E}^t = -\frac{1-\varkappa ^2-\sigma ^2+ V_1}{2w^2},
\end{equation}

It should be noted that the magnetization state in the helix nanowire was recently studied:\cite{Tkachenko13} in particular, the magnon spectrum was shown to be affected by the curvature, which acts mainly as effective anisotropy. However the equilibrium state was forcedly supposed to be the tangential one in Ref.~\onlinecite{Tkachenko13}.

\subsection{Onion state}
\label{sec:onion}

Let us discuss the case of a large curvature and torsion. In analogy with the ring wire, we are looking for a solution periodic with respect to $\chi$, which is an analogue of the onion solution \eqref{eq:vortex-onion}. Hence we look for solutions of the following form
\begin{subequations} \label{eq:onion}
\begin{equation} \label{eq:onion-1}
\theta ^{\mathrm{on}}(s) = \frac{\pi }{2} + \vartheta (\chi ), \quad \phi ^{\mathrm{on}}(s) = -\chi  + \varphi (\chi )
\end{equation}
with $\vartheta (\chi )$ and $\varphi (\chi )$ being $2\pi $--periodic functions. Using an analogy with the ring case ($\sigma =0$) with exact onion solution \eqref{eq:vortex-onion} we name \eqref{eq:onion-1} an \emph{onion} solution.

Numerically we found onion solutions for $\varkappa >\varkappa _0\approx 0.657$ in a wide range of $\sigma $, see Figs.~\ref{fig:helix}(c), \ref{fig:helix_phd}(a). 
The symmetry of the static form of Eqs.~\eqref{eq:LLE-helix} dictates the symmetry of $2\pi$--periodic functions $\vartheta$ and $\varphi$, which has the following  Fourier expansion
\begin{equation} \label{eq:onion-2}
\vartheta (\chi ) =  \sum_{n=1}^N \vartheta _n \cos (2n-1)\chi ,\quad \varphi (\chi ) = \sum_{n=1}^N \varphi _n \sin 2n \chi,
\end{equation}
\end{subequations}
where $N\to\infty$. By substituting series \eqref{eq:onion-2} into the static version of Eqs.~\eqref{eq:LLE-helix}, one get the set of nonlinear equations for amplitudes $\vartheta _n$ and $\varphi _n$, see \eqref{eq:LL-F-G}. Finally, the energy of the onion state $\mathscr{E}^{\mathrm{on}}(\sigma ,\varkappa _b)$, averaged over the helix period, can be calculated numerically using amplitudes $\vartheta _n$ and $\varphi _n$, see Appendix~\ref{app:onion} for details.

\subsection{Phase diagram}
\label{sec:phase-diagram}

%==================================================================\
\begin{figure*}
\begin{tikzpicture}%[scale=1,show grid={left,above,true}]
\node[right] at (0,-0.10) {\includegraphics[height=0.28\textwidth]{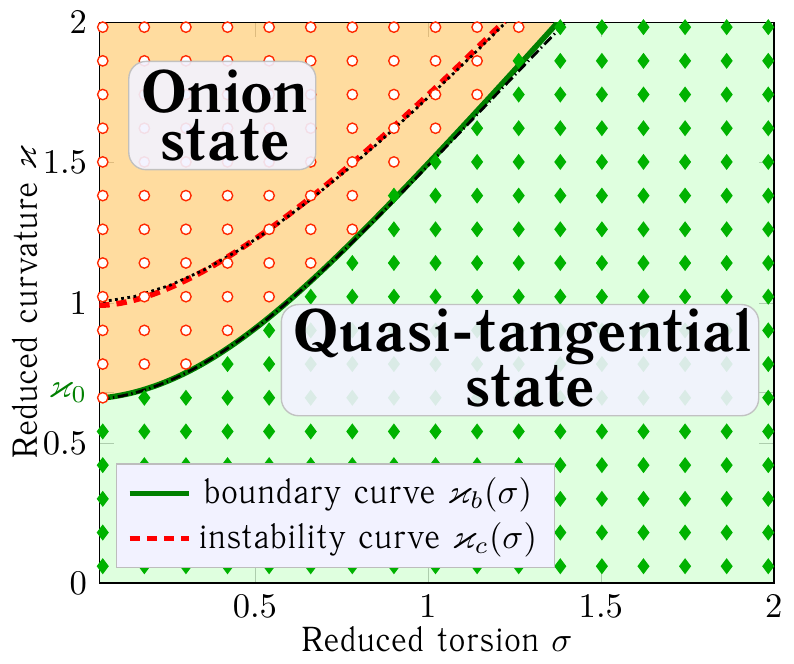}};
\node[right] at (5.9,-0.18) {\includegraphics[height=0.271\textwidth]{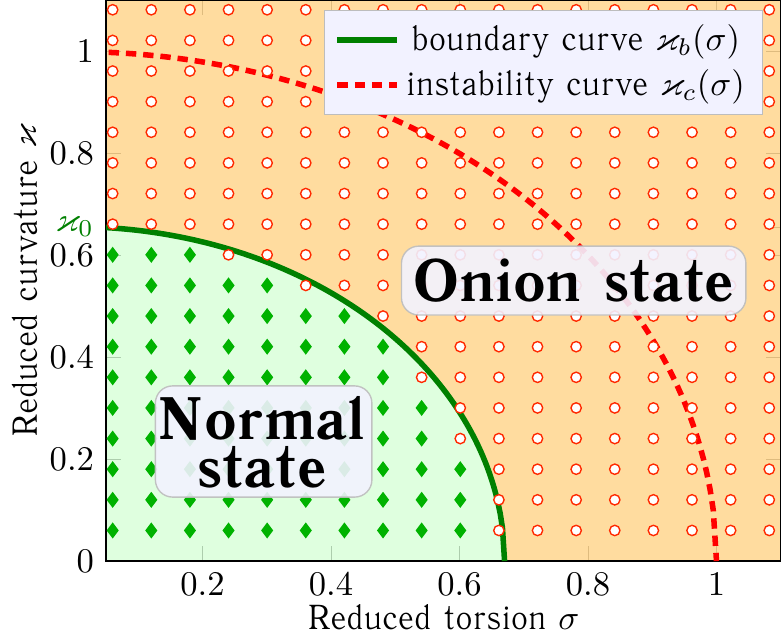}};
\node[right] at (11.9,-0.10) {\includegraphics[height=0.28\textwidth]{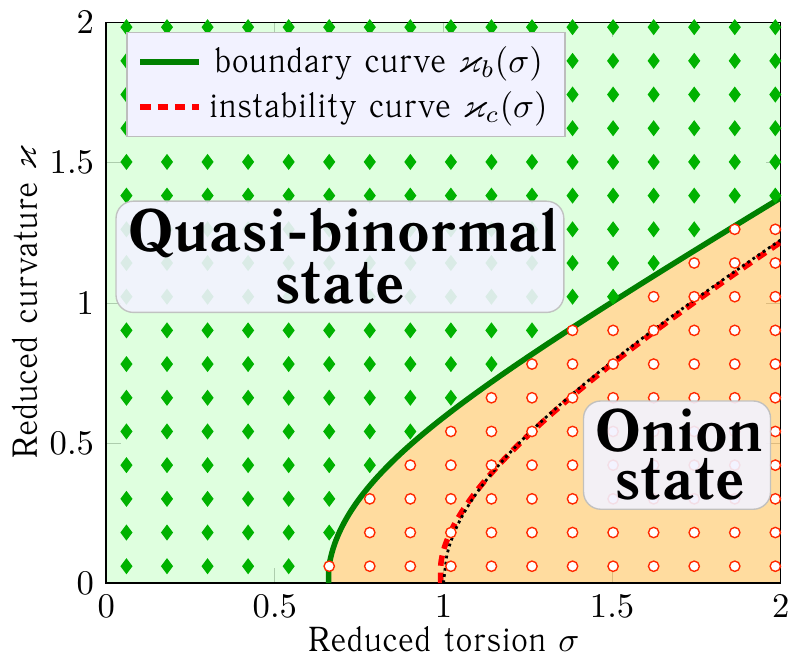}};
\node[right] at (1.2,-2.7) {(a) Easy-tangential anisotropy};
\node[right] at (7.2,-2.7) {(b) Easy-normal anisotropy};
\node[right] at (13,-2.7) {(c) Easy-binormal anisotropy};
\end{tikzpicture}
\caption{(Color online) Phase diagram of equilibrium magnetization states for the helix wire with different types of anisotropy. Symbols correspond to simulation data: green diamonds to homogeneous (in curvilinear reference frame) states and open circles to the onion ones. (a) Easy--tangential case, the curve $\varkappa_b(\sigma )$ (solid green line), calculated by \eqref{eq:boundary-condition} with $N=3$, describes the boundary between the quasi-tangential and the onion states; the dashed-dot line corresponds to $\varkappa_b(\sigma )$ with $N=1$. The curve $\varkappa _c(\sigma )$ (dashed red line) describes the boundary of linear instability of the quasi-tangential state, the dotted line is its fitting by \eqref{eq:sigma-kappa}. In the region between lines $\varkappa _b(\sigma )$ and $\varkappa _c(\sigma )$ the quasi-tangential state is metastable. (b) and (c) correspond to easy-normal and easy-binormal anisotropy, respectively; all notations have the same sense as in (a). Note that Fig.~(b) has different scale in order to show the normal state region in details.}
\label{fig:helix_phd}
\end{figure*}
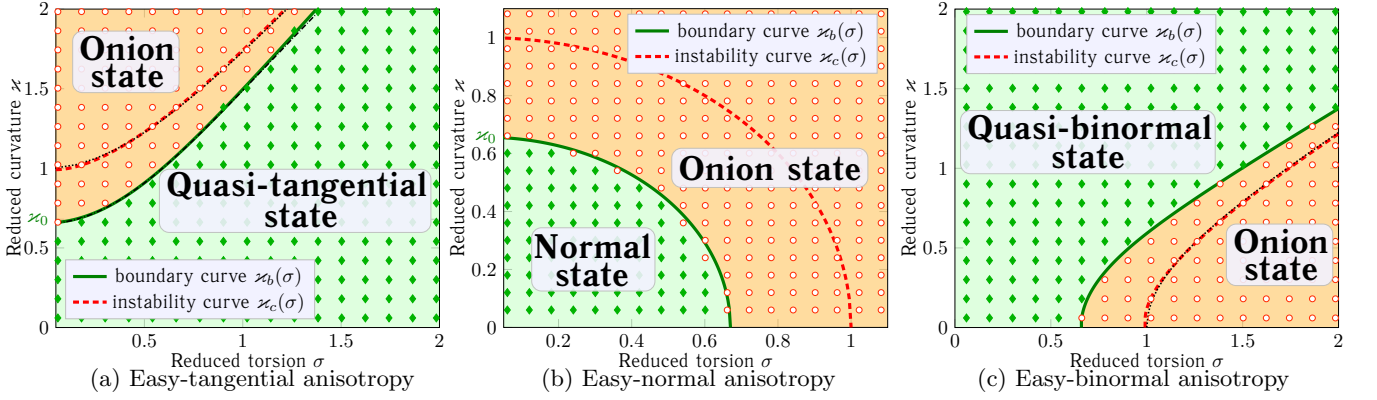
%==================================================================/
%

Now we summarize results on the equilibrium magnetization distribution. By comparing energies of different states, we compute the energetically preferable states for different curvature and torsion values. The resulting phase diagram is presented in Fig.~\ref{fig:helix_phd}(a). There are two phases: (i) The quasi-tangential state is realized for relatively small curvatures, when $\varkappa <\varkappa _b(\sigma )$; in such a state the magnetization direction is close to the direction of easy--tangential anisotropy $\vec{e}_{\text{\sc{t}}}$, see Fig.~\ref{fig:helix}(a,b) with the limit vortex orientation in case of the ring wire ($\tau =0$). (ii) The onion state corresponds to the case, when $\varkappa >\varkappa _b(\sigma )$; the magnetization distribution is inhomogeneous in accordance to \eqref{eq:onion}, see Fig.~\ref{fig:helix}(c).

The boundary between two phases $\varkappa _b=\varkappa _b(\sigma )$ can be derived using the condition
\begin{equation} \label{eq:boundary-condition}
\mathscr{E}^t(\sigma ,\varkappa _b) = \mathscr{E}^{\mathrm{on}}(\sigma ,\varkappa _b),
\end{equation}
where ${\mathscr{E}}^{\mathrm{on}}$ is energy density of the onion state averaged over the helix period $2\pi s_0$, see \eqref{eq:En-onion}. The onion solution \eqref{eq:onion} is energetically preferable when its energy is lower than the energy of the quasi-tangential state \eqref{eq:helix-gs-energy}. We computed the boundary curve numerically for $N=1$ and $N=3$, see dot-dashed and solid lines, respectively in the Fig.~\ref{fig:helix_phd}(a). The obtained curves are very close, so the approximation $N=1$ is reasonable. This is because the onion state of the helix wire is very close to an uniform magnetization, see Fig.~\ref{fig:helix}(c).

For the approximate description of the boundary dependence we use the trial function
\begin{equation} \label{eq:phd-trial}
\varkappa _b^{\mathrm{ET}} = \sqrt{\varkappa _0^2+2\sigma^2}, 
\end{equation}
which fits the numerically calculated curve $\varkappa _b(\sigma )$ with an accuracy of about $5\times 10^{-2}$.

Using the boundary dependence $\varkappa _b(\sigma )$, one can easily compute domain of applicability of the quasi-tangential solution \eqref{eq:helix-gs-1}:
\begin{subequations} \label{eq:theta-b}
\begin{equation} \label{eq:theta-b-ineq}
\theta^t \in
\begin{cases}
\left(\theta_b^t,\frac{\pi}{2}\right), & \text{when $\mathcal{C}\sigma>0$},\\
\left(\frac{\pi}{2},\theta_b^t\right), & \text{when $\mathcal{C}\sigma<0$}.
\end{cases}
\end{equation}
Here $\theta_b^t = \theta_b^t(\sigma)$ determines the boundary curve,
\begin{equation} \label{eq:theta-b-curve}
\theta_b^t(\sigma) \equiv \theta^t(\varkappa_b(\sigma),\sigma).
\end{equation}
\end{subequations}

%\begin{equation} \label{eq:phd-trial}
%\begin{split}
%\varkappa _b^{\mathrm{ET}} &= \sqrt{\varkappa _0^2+2c_1(\sigma -\sigma _0)^2}, \\
%c_1&\approx 1.04, \qquad \sigma _0\approx 0.06,
%\end{split}
%\end{equation}
%which fits numerically calculated curve $\varkappa _b(\sigma )$ with an accuracy of about $5\times 10^{-3}$.

\section{Spin wave spectrum in a helix wire with easy-tangential anisotropy}
\label{sec:sw}

We limit our consideration of spin waves by the case of the quasi-tangential magnetization state. First we linearize the Landau--Lifshitz equations \eqref{eq:LLE-helix} on the background of the quasi-tangential equilibrium state \eqref{eq:helix-gs-1},
\begin{equation} \label{eq:vartheta-vatphi}
\theta (s,t) = \theta^t + \vartheta (s,t), \qquad \phi (s,t) = \phi ^t + \frac{\varphi (s,t)}{\sin\theta ^t}.
\end{equation}
Then  for $\vartheta $ and $\varphi $ we get the set of linear equations:
\begin{equation} \label{eq:linear}
\begin{split}
\partial_{t'} \varphi   &=   -\partial_{\xi \xi }\vartheta  + V_1 \vartheta  - 2 A \partial_\xi  \varphi ,\\
-\partial_{t'} \vartheta  &=   -\partial_{\xi \xi }\varphi  + V_2 \varphi  + 2 A \partial_\xi  \vartheta ,
\end{split}
\end{equation}
where $\partial_{t'}$ is the derivative with respect to dimensionless time $t'=\Omega _0 t$ with $\Omega _0=2K\gamma/M_s$ and $\partial_\xi $ is the derivative with respect to dimensionless coordinate $\xi =s/w$. Here $V_1$ is determined according to \eqref{eq:helix-gs-1}, the quantities $V_2$ and $A$ have the following form:
\begin{equation} \label{eq:V2-A}
\begin{split}
V_2 &= \frac{1+\varkappa ^2+\sigma ^2+V_1}{2},\\
A &= -\varkappa \cos\theta ^t - \sigma  \mathcal{C}\sin\theta ^t = -\sigma  \mathcal{C}V_2\sqrt{\frac{2}{V_1 V_0}}.
\end{split}
\end{equation}
While $V_1$ and $V_2$ appear as scalar potentials, $A$ acts as a vector potential $\vec{A} = A\vec{e}_{\text{\sc{t}}}$ of effective magnetic field. This becomes obvious if we combine the set of linearized equations for $\vartheta $ and $\varphi $ in a single equation for the complex-valued function
$\psi =\vartheta +i\varphi $,
\begin{subequations} \label{eq:Schroedinger}
\begin{equation} \label{eq:Schroedinger-ET}
-i\partial_{t'} \psi  = H \psi  + W\psi ^*, \quad H = \left(-i\partial_\xi  -A \right)^2 +U.
\end{equation}
This differential equation has a form of generalized Scr{\"o}dinger equation, originally proposed for the description of spin waves on the magnetic vortex background.\cite{Sheka04} The ``potentials'' in Eq.~\eqref{eq:Schroedinger-ET} read
\begin{equation} \label{eq:potentials}
U = \frac{V_1+V_2}{2}-A^2, \quad W = \frac{V_1-V_2}{2} = -\frac{1+w^2\mathscr{E}^t}{2}.
\end{equation}
\end{subequations}
An effective magnetic field $\vec{A}$ is originated from the curvature induced effective Dzyaloshinskii interaction, see Eq.~\eqref{eq:energy-Frenet}: the energy density $\mathscr{E}_{\mathrm{ex}}^{D}$, harmonized using \eqref{eq:vartheta-vatphi}, reads\footnote{Up to a full derivative with respect to $\xi$.}
\begin{equation} \label{eq:A-v-a-ED}
\mathscr{E}_{\mathrm{ex}}^{D} = -\frac{2}{w^2}A |\psi |^2\partial_\xi  \arg \psi .
\end{equation}
Now we apply the traveling wave Ansatz for the spin-wave complex magnon amplitude
\begin{equation} \label{eq:spin-wave}
\psi (\xi ,t' ) = \mathrm{u} e^{i\Phi } + \mathrm{v} e^{-i\Phi },\qquad \Phi  = q\xi -\Omega t' +\eta,
\end{equation}
with $q=k w$ being the dimensionless wave number, $\Omega =\omega /\Omega _0$ the dimensionless frequency, $\eta $ is arbitrary phase, and $ \mathrm{u}, \mathrm{v}\in\mathbb{R}$ being constants. The corresponding wave vector is oriented along the wire, $\vec{q} = q\vec{e}_{\text{\sc{t}}}$; its orientation with respect to the equilibrium magnetization is determined by Eq.~\eqref{eq:helix-gs-1}. By substituting the Ansatz \eqref{eq:spin-wave} into the generalized Scr{\"o}dinger equation \eqref{eq:Schroedinger}, one can derive the spectrum of the spin waves:
\begin{equation} \label{eq:disp}
\Omega (q) = 2Aq+ \sqrt{\left(q^2+V_1\right)\left(q^2+V_2\right)}.
\end{equation}

\begin{figure*}
\includegraphics[width=\textwidth]{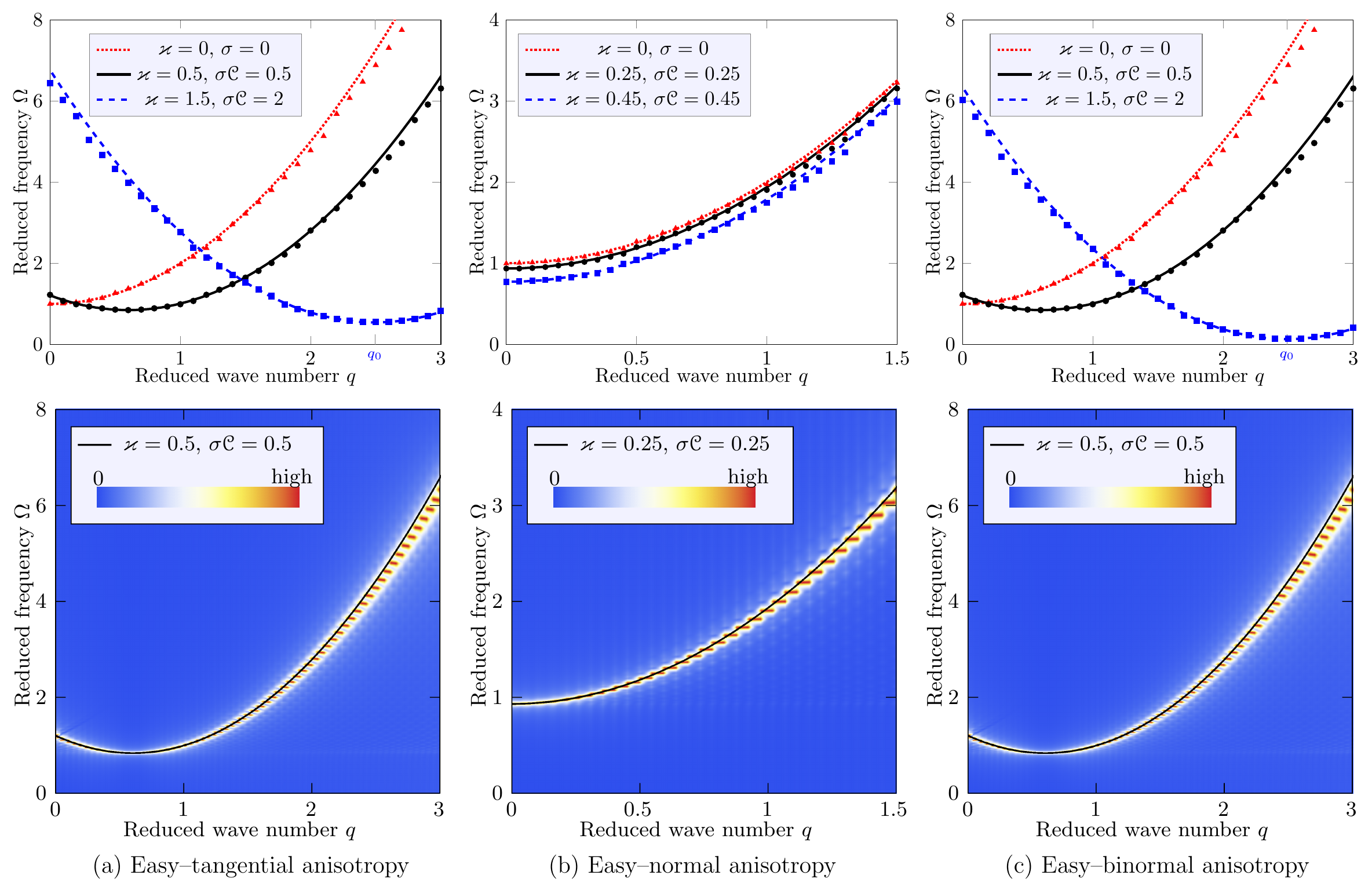}
\caption{(Color online) Top row demonstrates dispersion laws for spin waves in the helix wire for different anisotropies. The equilibrium states are homogeneous in the curvilinear reference frame. Symbols correspond to simulation data, see Sec.~\ref{sec:spinwaves-simulations}, and lines to the analytics, see Eq.~\eqref{eq:disp} and \eqref{eq:disp-EN}. Few examples of dispersion relation are shown at the bottom row in terms of density plots to demonstrate that \eqref{eq:disp} is a single frequency branch in the system.
}
\label{fig:helix_dispersion}
\end{figure*}

Similar to the straight wire case with $\Omega _s(q) = 1+q^2$, the spectrum of spin waves in the helix wire has a gap, caused, first of all, by the anisotropy (in dimensional units the gap has an order of $\Omega_0\propto K$). However its value essentially depends on the curvature and the torsion. Moreover, the spectrum gap occurs at finite $q=q_0$, see Fig.~\ref{fig:helix_dispersion}. This means the asymmetry in the spectrum with respect to the change $q\to-q$: spin waves have different velocities depending on the direction (along the helix axis or in opposite direction). This asymmetry in the dispersion law \eqref{eq:disp} occurs in the first term $2Aq$, which is originated from the effective Dzyaloshinskii interaction $\mathscr{E}_{\mathrm{ex}}^{D}$.

In this context it is instructive to mention that the spin wave spectrum in the presence of Dzyaloshinskii-Moriya interaction is known to be asymmetric with respect to wave vector inversion and has the minimum at finite wave vectors. \cite{Zakeri10,Cortes-Ortuno13,Zakeri14} The curvature induced asymmetry in the spin waves propagation in nanotubes and its analogy with the Dzyaloshinskii-Moriya interaction was discussed recently in Ref.~\onlinecite{Hertel13a}. The spin-wave spectrum for the helix wire was calculated recently in Ref.~\onlinecite{Tkachenko13}, however the deviations from the pure tangential state were no taken into account and the effective Dzyaloshinskii was not considered.

In order to make analytical estimations, we consider now the dispersion law in case of very small curvatures and torsions:
\begin{equation} \label{eq:disp4wto0}
\begin{split}
\Omega (q) &=  \Omega_{\mathrm{gap}} +\left(q- \mathcal{C} \sigma \right)^2 + {\scriptstyle\mathcal{O}}\left(\varkappa^2, \sigma^2, \varkappa \sigma \right),\\
\Omega_{\mathrm{gap}} &= 1-\frac{\varkappa ^2}{2}+ {\scriptstyle\mathcal{O}}\left(\varkappa^2, \sigma^2, \varkappa \sigma \right).
\end{split}
\end{equation}
One can see that the spin wave spectrum becomes asymmetrical one with increasing the curvature and the torsion: the minimum of the frequency corresponds to $q_0=\sigma \mathcal{C}$ (in dimensional units the corresponding wave number $k_0 = \tau \mathcal{C}$), its sign is determined by the product of the helix chirality and the magnetochirality.

%==================================================================\
%==================================================================/

The further increase of the curvature and torsion decrease the gap $\Omega_{\mathrm{gap}}$; there is a critical curve $\varkappa _c = \varkappa _c(\sigma )$, where the gap vanishes, $\Omega (q_c)=0$ and $\partial_q \Omega (q_c)=0$. One can easily find that $q_c=\mathcal{C}\sqrt{A^2-U}$ and the critical curve $\varkappa _c = \varkappa _c(\sigma )$ can be found as a solution of algebraic equation
\begin{equation} \label{eq:kappa-critical}
4A^2U = W^2.
\end{equation}
The critical curve $\varkappa _c(\sigma )$, calculated numerically is plotted in Fig.~\ref{fig:helix_phd}(a) (dashed red curve). For the approximate description of the critical dependence we use the trial function
\begin{equation} \label{eq:sigma-kappa}
\varkappa _c^{\mathrm{trial}} = \sqrt{1+2\sigma ^2},
\end{equation}
which fits the numerical results with an accuracy of about $5\times10^{-3}$, see the dotted curve in Fig.~\ref{fig:helix_phd}(a). In the region between the boundary curve $\varkappa _b(\sigma )$ and the instability curve $\varkappa _c(\sigma )$ [see Fig.~\ref{fig:helix_phd}] the quasi-tangential state becomes metastable.

\section{Helix with other anisotropy orientations}
\label{sec:anis}

Let us discuss other types of anisotropies: easy--normal and easy--binormal, see Eq.~\eqref{eq:Ean-angular} and Table~\ref{tab:anisotropy}.

\subsection{Easy--normal anisotropy}
\label{sec:EN}

Let us start the analysis of the easy--normal anisotropy with the limit case of the ring ($\tau =0$). In this case, similarly to the easy--tangential anisotropy, the magnetization lies within the ring plane: $\theta =\pi /2$. The energy minimization with respect to $\phi $ results in the pendulum equation:
\begin{equation} \label{eq:pendulum-EN}
\varkappa ^2 \partial_{\chi \chi }\phi  +\sin\phi \cos\phi =0.
\end{equation}
In analogy with the easy--tangential anisotropy the equilibrium state is the exactly normal state $\phi^n=\pm\pi/2$ in case of relatively small reduced curvature $\varkappa <\varkappa _0$ and inhomogeneous onion solution $\phi^{\mathrm{on}}_n(\chi)=\pi/2-\phi^{\mathrm{on}}(\chi)$ for $\varkappa >\varkappa _0$, where function $\phi^{\mathrm{on}}(\chi)$ is defined by \eqref{eq:vortex-onion}.

\begin{figure}
\begin{center}
\includegraphics[width=\columnwidth]{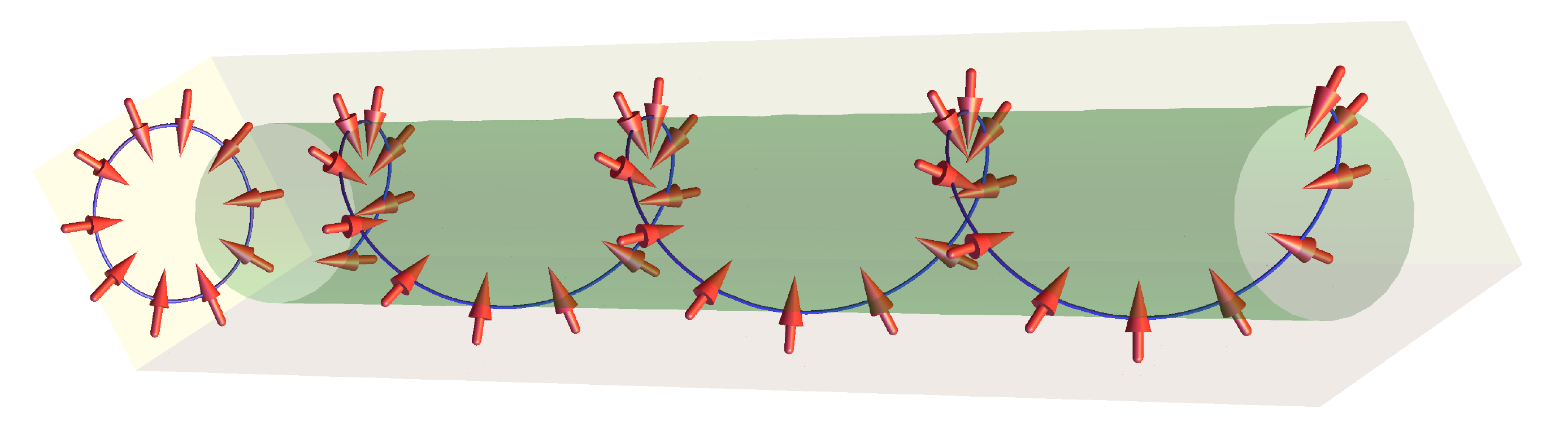}\\
(a) Normal state: $\varkappa =\sigma =0.45$\\
\includegraphics[width=\columnwidth]{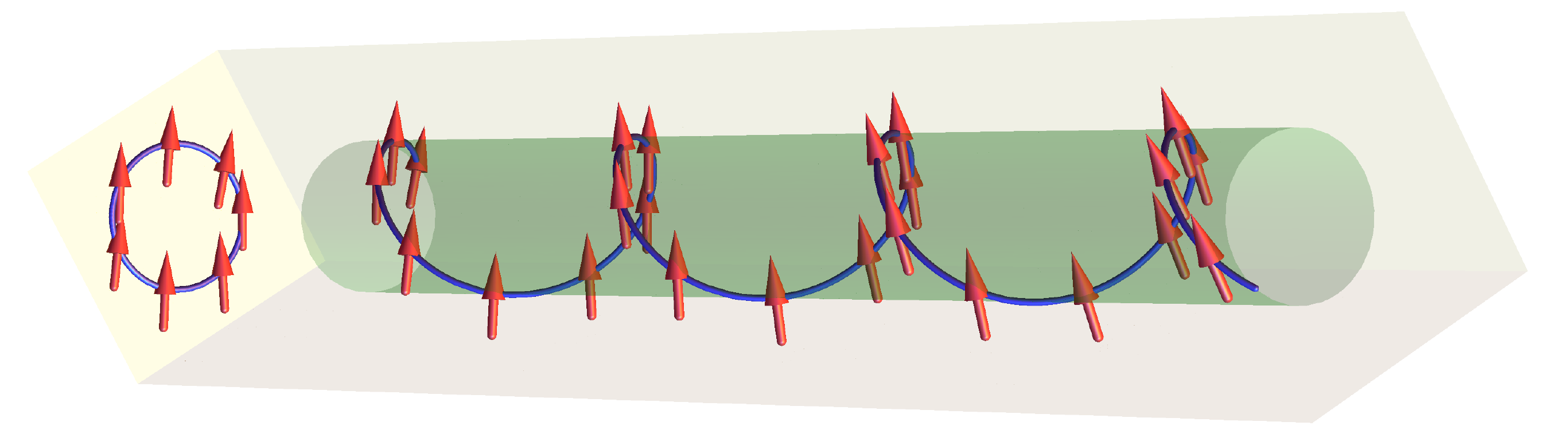}\\
(b) Onion state: $\varkappa =\sigma =1$\\
\end{center}
\caption{(Color online) Magnetization distribution in the helix wire with $\mathcal{C}=+1$ and easy-normal anisotropy according to simulations data, see Sec.~\ref{sec:equilibrium-simulations}.}
\label{fig:normal}
\end{figure}

In case of finite torsion there also exists \emph{exactly normal} state
\begin{equation} \label{eq:helix-gs-radial}
\theta ^n =\frac{\pi }{2}, \quad \phi ^n = \mathcal{C}\frac{\pi }{2}, \qquad \mathscr{E}^n = -\frac{1-\varkappa ^2-\sigma ^2}{w^2},
\end{equation}
where $\mathcal{C}=\pm1$, see Fig.~\ref{fig:normal}(a). Such a state is energetically preferable for relatively small values of $\varkappa $ and $\sigma $. The magnetization in the normal state is directed exactly radially, which is well pronounced in experiments with 3D microhelix coil strips.\cite{Smith11}

%==============================================================================/

%==============================================================================/

In case of large curvature, there is the periodic (in curvilinear reference frame) onion solution, which has the form \eqref{eq:onion}, see Fig.~\ref{fig:normal}(b). Using the same numerical procedure as in Sec.~\ref{sec:onion}, we evaluate the onion solution and compute the phase diagram, see Fig.~\ref{fig:helix_phd}(b).

For the approximate description of the boundary $\varkappa _b^{\mathrm{EN}}(\sigma)$ between two phases we use the fitting function
\begin{equation} \label{eq:phd-trial-EN}
\varkappa _b^{\mathrm{EN}} = \varkappa _0\sqrt{1- \left(\frac{\sigma }{\sigma _0}\right)^2}, \quad \sigma _0\approx 0.67,
\end{equation}
which fits the numerically calculated curve $\varkappa_b^{\mathrm{EN}}(\sigma )$ with an accuracy of about $3\times 10^{-3}$. 

Let us discuss now the linear excitations on the background of the normal solution. Using the same approach as in Sec.~\ref{sec:sw}, we linearize Landau--Lifshitz equations \eqref{eq:LLE-helix} on the background of the normal solution  \eqref{eq:helix-gs-radial}, $\theta =\theta ^n+\vartheta $, $\phi =\phi ^n+\varphi $. After linearization one gets a generalized Scr{\"o}dinger--like equation for the complex variable $\psi =\vartheta +i\varphi $,
\begin{subequations} \label{eq:Schroedinger-EN}
\begin{equation} \label{eq:Schroedinger-EN-1}
-i\partial_{t'} \psi  = \left(-\partial_{\xi \xi } +U^n\right) \psi  + W^n\psi ^*.
\end{equation}
Here the ``potentials'' read
\begin{equation} \label{eq:potentials-EN}
U^n = 1- \frac{\varkappa ^2+\sigma ^2}{2}, \qquad W^n = \frac12 \left(\mathcal{C}\sigma  - i\varkappa  \right)^2.
\end{equation}
\end{subequations}
Let us compare this equations with the generalized Scr{\"o}dinger--like equation \eqref{eq:Schroedinger}. First of all, there is no effective vector potential, since there is no asymmetry by effective Dzyaloshinskii interaction like in easy--tangential case. The second difference is that the potential $W$ in \eqref{eq:potentials-EN} is a complex--valued one, hence the scattering problem is similar to the two--channel scattering process. Similar to \eqref{eq:spin-wave} we apply the following traveling wave Ansatz for the spin-wave complex magnon amplitude
\begin{equation} \label{eq:spin-wave-EN}
\psi (\xi ,t' ) = \psi _1 e^{i\Phi } + \psi _2 e^{-i\Phi },\; \Phi  = q\xi -\Omega t' +\eta , \; \psi _{1,2}\in\mathbb{C}.
\end{equation}
The difference is that constants $\psi _{1,2}$ are complex ones. Now by substituting the Ansatz \eqref{eq:spin-wave-EN} into the generalized Scr{\"o}dinger equation \eqref{eq:Schroedinger-EN}, one can derive the spectrum of the spin waves:
\begin{equation} \label{eq:disp-EN}
\Omega (q) = \sqrt{\left(1+q^2\right)\left(1+q^2-\varkappa ^2-\sigma ^2\right)}.
\end{equation}
This dispersion relation is reproduced by the numerical simulations with a high accuracy, see Fig.~\ref{fig:helix_dispersion}(b). The critical dependence, where the gap of the spectrum vanishes, reads
\begin{equation} \label{eq:EN-instab}
\varkappa _c = \sqrt{1-\sigma ^2},
\end{equation}
see thick dashed curve in Fig.~\ref{fig:helix_dispersion}(b). In the region between solid and dashed curves the normal state is metastable.

The dispersion law \eqref{eq:disp-EN} is symmetric with respect to the direction of the wave propagation: $\Omega (q) = \Omega (-q)$. Unlike the easy-tangential case there is no effective magnetic field $\vec{A}$, because the curvature induced effective Dzyaloshinskii interaction is absent in the harmonic approximation, cf.~\eqref{eq:A-v-a-ED}. The reason is that the equilibrium state is magnetized exactly in the normal direction $\vec{e}_{\text{\sc{n}}}$, which causes the degeneracy with respect to the sign of $q$. A similar behavior is known for thin films in the presence of Dzyaloshinskii--Moriya interaction, where the asymmetry in the spin wave spectrum vanishes if the system is saturated perpendicularly to the film plane.\cite{Cortes-Ortuno13}

\subsection{Easy--binormal anisotropy}
\label{sec:EB}

%==============================================================================/
\begin{figure}
\begin{center}
\includegraphics[width=\columnwidth]{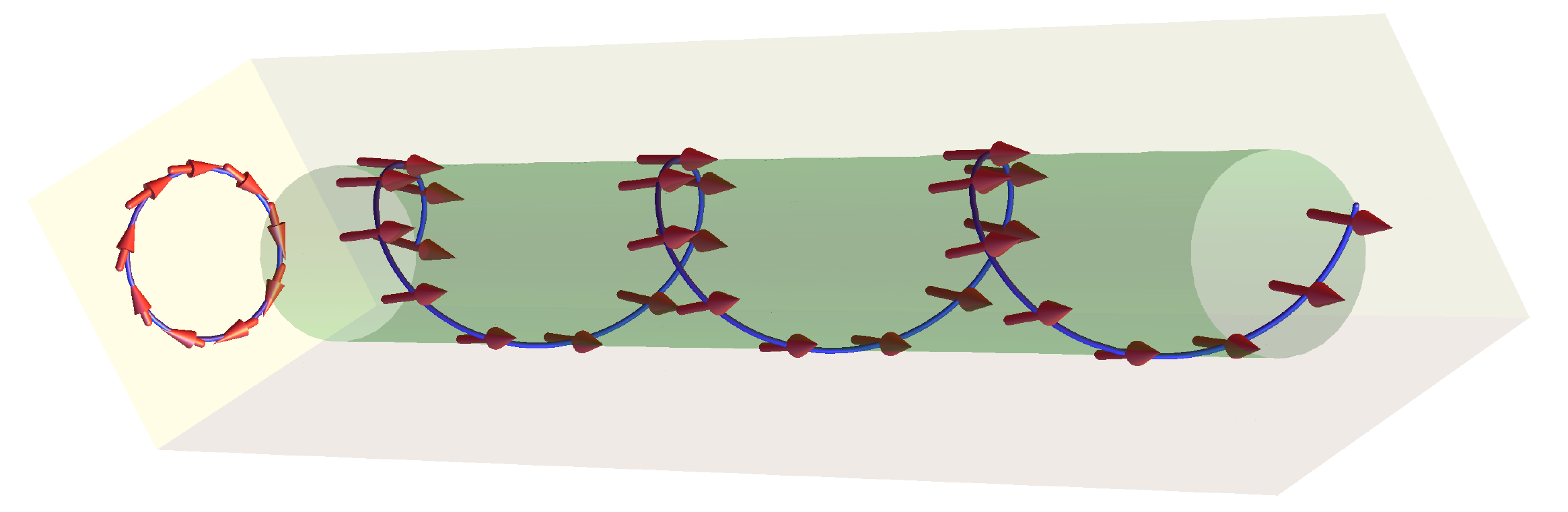}\\
(a) Quasi-binormal state: $\varkappa =\sigma =1.5$\\
\includegraphics[width=\columnwidth]{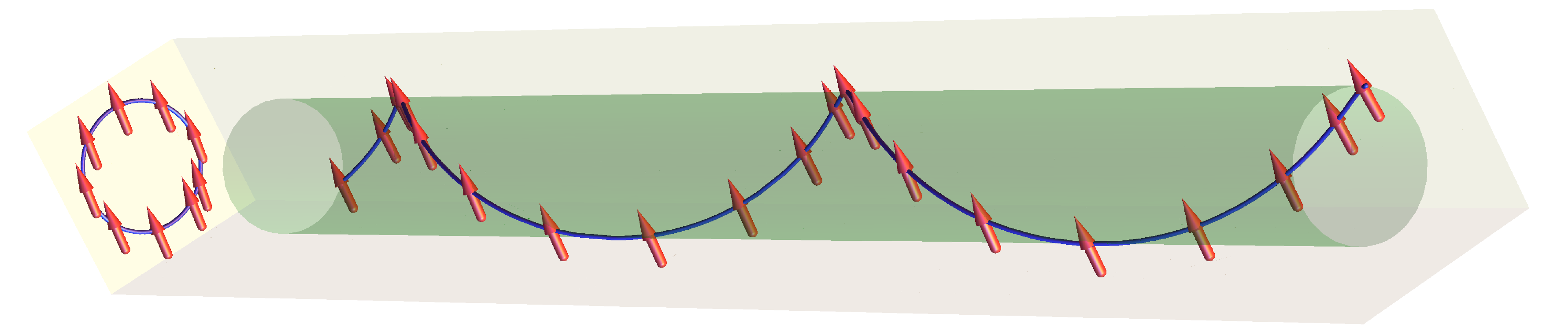}\\
(b) Onion state: $\varkappa =1,$ $\sigma =2$\\
\end{center}
\caption{(Color online) Magnetization distribution in the helix wire with $\mathcal{C}=+1$ and easy-binormal anisotropy according to simulations data, see Sec.~\ref{sec:equilibrium-simulations}.}
\label{fig:binormal}
\end{figure}
%==============================================================================/

If the anisotropy axis is directed along $\vec{e}_{\text{\sc{b}}}$, one has the easy--binormal anisotropy, $\mathscr{E}_{\text{an}}^{\mathrm{EB}}$, see \eqref{eq:Ean-angular}. The magnetization of the homogeneous (in the curvilinear reference frame) state reads
\begin{equation} \label{eq:helix-gs-binormal}
\tan 2 \theta ^b = \frac{2 \mathcal{C} \varkappa \sigma }{1+\varkappa ^2-\sigma ^2}, \qquad \cos \phi ^b=\mathcal{C}=\pm1,
\end{equation}
Explicitly $\theta ^b$ reads
\begin{equation} \label{eq:helix-gs-EB}
\begin{split}
& \theta ^b = \frac{\pi }{2}\left[1+\mathrm{sgn}\left(\mathcal{C}\sigma \right) \right] -  \arctan\frac{2\mathcal{C}\sigma \varkappa }{V_0^b},\\
& V_0^b = 1+\varkappa ^2-\sigma ^2+V_1^b,\\
& V_1^b = \sqrt{(1+\varkappa ^2-\sigma ^2)^2+4\varkappa ^2\sigma ^2}.
\end{split}
\end{equation}
The magnetization of this state is close to the direction of the helix axis, hence we name it \emph{quasi--binormal} state, see Fig.~\ref{fig:binormal}(a). It corresponds to the hollow--bar magnetization distribution in the helix microcoils.\cite{Smith11}  For different magnetization distributions see also Table~\ref{tab:anisotropy}.

The energy of the axial state reads
\begin{equation} \label{eq:helix-gs-energy-binormal}
\mathscr{E}^b = -\frac{1-\varkappa ^2-\sigma ^2+ V_1^b}{2w^2}.
\end{equation}
Let us mention the formal analogy between the energy $\mathscr{E}^b$, the ``potentials'' $V_0^b$, $V_1^b$ for the quasi-binormal state and the corresponding expressions $\mathscr{E}^t$ [cf. \eqref{eq:helix-gs-energy}], $V_0^t$, $V_1^t$ [cf. \eqref{eq:helix-gs-1}] for the quasi-tangential state: the expressions for the quasi-tangential state can be used for the quasi-binormal one under the replacement $\varkappa \leftrightarrow \sigma $.

The analogy between two states becomes deeper if we use another parametrization for the magnetization $\vec{m}$
\begin{equation} \label{eq:theta-phi-1D-another}
\vec{m} = \cos\Theta \, \vec{e}_{\text{\sc{t}}} - \sin\Theta \sin\Phi \, \vec{e}_{\text{\sc{n}}} + \sin\Theta \cos\Phi \,\vec{e}_{\text{\sc{b}}},
\end{equation}
where $\Theta =\Theta (s)$ and $\Phi =\Phi (s)$ are the angles in the Frenet--Serret frame of reference: the polar angle $\Theta $ describes the deviation of magnetization from the tangential curve direction, while the azimuthal angle $\Phi $ corresponds to the deviation from the binormal. Similar to \eqref{eq:energy-angular}, one can rewrite the energy terms as follows (cf. Appendix A from the Ref.~\onlinecite{Sheka15} for details):
\begin{equation} \label{eq:energy-angular-bi}
\begin{split}
&\mathscr{E}_{\text{ex}} = \left[\Theta '-\kappa \sin\Phi \right]^2 + \left[\sin\Theta  (\Phi '+\tau ) -  \kappa \cos\Theta \cos\Phi  \right]^2,\\
%\end{equation}
&\mathscr{E}_{\text{an}}^{\mathrm{EB}} = -\frac{\sin^2\Theta \cos^2\Phi }{w^2}.
\end{split}
\end{equation}
Now one can easily see that the energy functional of the easy--tangential magnet transforms to the energy functional of the easy--binormal magnet under the following conjugations: $\theta \rightarrow \Theta $, $\phi \rightarrow\Phi $, and $\varkappa \leftrightarrow \sigma $.

Similarly to the easy--tangential case, there exist two equilibrium states: the homogeneous state (quasi-binormal) and the periodic onion solution, see Fig.~\ref{fig:binormal}(b). The phase diagram, which separates these two states, is plotted in the Fig.~\ref{fig:helix_phd}(c).

Now we discuss the magnons for the easy--binormal case. In analogy with the easy--tangential case, the linearized equations can be reduced to the generalized Scr{\"o}dinger equation \eqref{eq:Schroedinger-ET} with the following ``potentials'':
\begin{equation} \label{eq:V2-A-EB}
\begin{split}
V_2^b &= \frac{1+\varkappa ^2+\sigma ^2+V_1^b}{2},\\
A^b &= -\varkappa \cos\theta ^b - \sigma  \mathcal{C}\sin\theta ^b = -\varkappa  \mathcal{C}V_2^b\sqrt{\frac{2}{V_1^b V_0^b}}.
\end{split}
\end{equation}
The dispersion law has formally the form \eqref{eq:disp} with the corresponding ``potentials'' described above. The dispersion curve is plotted in the Fig.~\ref{fig:helix_dispersion}(c) for some typical parameters, it is confirmed by the numerical simulations. The critical curve $\varkappa _c(\sigma )$, where the gap of the spectrum vanishes, can be found numerically using condition \eqref{eq:kappa-critical}. The critical curve $\varkappa _c(\sigma )$, calculated numerically is plotted in Fig.~\ref{fig:helix_phd}(c) (dashed red curve). For the approximate description of the critical dependence we use the trial function
\begin{equation} \label{eq:sigma-kappa-ET}
\varkappa _c^{\mathrm{trial}} = \sqrt{\frac{\sigma ^2-1}{2}},
\end{equation}
which fits the numerical results of Fig.~\ref{fig:helix_phd}(c) with an accuracy of about $2\times10^{-2}$, see the dotted curve in Fig.~\ref{fig:helix_phd}(c). In the region between solid and dashed curves the quasi-binormal state is metastable.

\section{Simulations}
\label{sec:simulations}

In order to verify our analytical results we numerically simulate the magnetization dynamics of a helix-shaped chain of discrete magnetic moments $\vec m_i$ with $i=\overline{1,N}$. The form of the chain is described by Eq.~\eqref{eq:helix-param-kappatau}. Magnetization dynamics of this system is determined by the set of Landau--Lifshitz equations
\begin{equation} \label{eq:LL-eq-mod}
\frac{1}{\omega_0}\frac{\mathrm{d}\vec{m}_i}{\mathrm{d}t}=\vec{m_i} \times \frac{\partial\mathcal{E}}{\partial\vec{m}_i}+\alpha \vec{m}_i \times \left[ \vec{m}_i \times \frac{\partial\mathcal{E}}{\partial\vec{m}_i }\right],
\end{equation}
where $\omega_0=4\pi\gamma M_s$, $\alpha$ is the damping coefficient, $\mathcal{E}$ is the dimensionless energy, normalized by $4\pi M_s^2\Delta s^3$  with $\Delta s$ being the sampling step of the natural parameter $s$. We consider four contributions to the energy of the system:
\begin{subequations} \label{eq:tot-energy-eq-mod}
\begin{equation} \label{eq:tot-energy-mod}
\mathcal{E} = \mathcal{E}^{\text{ex}} + \mathcal{E}^{\text{an}} + \mathcal{E}^{\text{f}}+ \mathcal{E}^{\text{d}}.
\end{equation}
The first term in Eq.~\eqref{eq:tot-energy-mod} is the exchange energy
\begin{equation} \label{eq:energy-exch-mod}
\mathcal{E}^{\text{ex}} = -2 \frac{\ell^2}{\Delta s^2} \sum\limits_{i=1}^{N-1} \vec{m}_i \cdot \vec{m}_{i+1}
\end{equation}
with $\ell = \sqrt{A/(4\pi M_s^2)}$ being the exchange length. The second term determines the uniaxial anisotropy contribution
\begin{equation} \label{eq:energy-anis-mod}
\mathcal{E}^{\text{an}} =-\frac{Q}{2} \sum\limits_{i=1}^{N} (\vec{m}_i \cdot \vec{e}_i^{\text{an}})^2,
\end{equation}
where $\vec{e}_i^{\text{an}}$ is the coordinate dependent unit vector along the anisotropy axis and $Q$ is the quality factor, see \eqref{eq:Q-factor}. The third term determines interaction with the external magnetic field $\vec{b}$
\begin{equation} \label{eq:energy-ext-mod}
\mathcal{E}^{\text{f}} =-\sum\limits_{i=1}^{N}\vec{b}_i\cdot\vec{m}_i,
\end{equation}
where $\vec{b}_i$ is the dimensionless external field, normalized by $4\pi M_s$. 

The last term in \eqref{eq:tot-energy-mod} determines the dipolar interaction
\begin{equation} \label{eq:energy-dip-mod}
\mathcal{E}^{\text{d}} =\frac{(\Delta s)^3\!\!}{8\pi}  \sideset{}{'}\sum_{i,j=1}^N \frac{\vec{m}_i\cdot\vec{m}_j}{|\vec{r}_{ij}|^3}-3\frac{\left(\vec{m}_i\cdot \vec{r}_{ij} \right) \left(\vec{m}_j\cdot\vec{r}_{ij} \right)}{|\vec{r}_{ij}|^5}.
\end{equation}
where $\vec{r}_{ij}\equiv \vec{\gamma}_i - \vec{\gamma}_j$.
\end{subequations}

The dynamical problem is considered as a set of $3N$ ordinary differential equations \eqref{eq:LL-eq-mod} with respect to $3N$ unknown functions $m_i^{\text{x}}(t),\,m_i^{\text{y}}(t),\,m_i^{\text{z}}(t)$ with $i=\overline{1,N}$. For a given initial conditions the set \eqref{eq:LL-eq-mod} is integrated numerically. During the integration process the condition $|\vec{m}_i(t)|=1$ is controlled.

%The simulator solve a $3N$ of ordinary differential equations for $\vec{m}_i(\tau')=(m_i^{\text{x}},m_i^{\text{y}},m_i^{\text{z}})$ with $i=\overline{1N}$ and $N$ being a number of spins. During the calculations we check next condition on solutions $|\vec{m}_i(\tau')|=1$.

We considered the helix wire with length $L=500\Delta s$, the exchange length $\ell=3\Delta s$ and the quality factor $Q=2$ are fixed. The curvature $\kappa$ and the torsion $\tau$ were varied under the restriction $\kappa\Delta s/\pi\ll1$.

In most of simulations we neglect magnetic dipolar interaction and consider the Heisenberg magnet with the energy
\begin{equation} \label{eq:tot-energy-Heisenberg}
\mathcal{E}^{\text{H}} = \mathcal{E}^{\text{ex}} + \mathcal{E}^{\text{an}} + \mathcal{E}^{\text{f}}.
\end{equation}

\subsection{Equilibrium magnetization states}
\label{sec:equilibrium-simulations}

We start our simulations with easy--tangential magnets. In Sec.~\ref{sec:helix}	we found that the curvature and the torsion causes the deviation of the magnetization from the anisotropy direction, which results in the magnetization distribution \eqref{eq:helix-gs-1}; such results are presented in Fig.~\ref{fig:helix_gs} by the curves for three different value of the reduced curvature $\varkappa=0.1,\varkappa_c,1$ in the wide range of the torsion $\sigma\in(0;2)$. In order to verify our theoretical predictions we simulate numerically Landau--Lifshitz equations \eqref{eq:LL-eq-mod} in overdamped regime ($\alpha=0.1$) during a long time interval $\Delta t\gg(\alpha\omega_0)^{-1}$. 

Numerically we model the anisotropic Heisenberg magnet with the energy \eqref{eq:tot-energy-Heisenberg} and $Q=2$. Simulation data are presented in Fig.~\ref{fig:helix_gs} by filled symbols and labeled as (a); one can see an excellent agreement between out theory and simulations. The typical magnetization distribution is shown in Figs.~\ref{fig:helix}(a), (b) and (c) for the quasi-tangential states and in Figs.~\ref{fig:helix}(d) for the onion state.

We also perform simulations for other anisotropy types. The magnetization distribution for the helix wire with easy-normal anisotropy  is presented in Fig.~\ref{fig:normal}(a) for the normal state and Fig.~\ref{fig:normal}(b) for the onion one. For the case of easy-binormal anisotropy one has two possible states: the quasi-binormal one [see Fig.~\ref{fig:binormal}(a)] and the onion one [see Fig.~\ref{fig:binormal}(b)].

The second stage of our simulations is to find the equilibrium magnetization state of a given helix wire. Numerically we simulate  Eqs.~\eqref{eq:LL-eq-mod} as described above for five different initial states, namely the tangential, onion, normal, binormal, and the random states. The final static state with the lowest energy is considered to be the equilibrium magnetization state. We obtain that for each type of anisotropy the equilibrium state is either onion one or anisotropy-aligned state (quasi-tangential, normal and quasi-binormal state for easy-tangential, easy-normal and easy-binormal anisotropy, respectively). We present simulations data in Fig.~\ref{fig:helix_phd} by symbols together with theoretical results (plotted by lines). One can see a very good agreement between simulations and analytics.

\subsection{Dispersion relations}
\label{sec:spinwaves-simulations}

For each anisotropy-aligned equilibrium state the magnon dispersion relation is obtained numerically. It is carried out in two steps. In the first step the helix wire is relaxed in external spatially nonuniform weak magnetic field
\begin{equation} \notag
\vec{b}_{i}^j=b_0\vec{e}_i^{\text{d}}\cos s_ik^j
\end{equation}
for a range of wave-vectors $k^j=j/(300\Delta s)$ with $j=\overline{0,300}$. Here $b_0\ll1$ is the field amplitude, $s_i=(i-1)\Delta s$ is position of the magnetic moment $\vec m_i$. The coordinate dependent unit vector $\vec{e}_i^{\text{d}}$ determines the magnetic field direction: $\vec{e}_i^{\text{d}}=\vec{e}_{\text{\sc{n}}}$ for the quasi-tangential state and $\vec{e}_i^{\text{d}} =\vec{e}_{\text{\sc{t}}}$ for normal and quasi-binormal states.

In the second step we switch off the magnetic field and simulate the magnetization dynamics with the damping value $\alpha=0.01$ close to natural one. Then the space-time Fourier transform is performed for one of the magnetization components (we consider normal component for the quasi-tangential state and tangential component for other two equilibrium states). The frequency $\Omega$ which corresponds to the maximum of the Fourier signal is marked by a symbol for a given wave-vector $q^j=wk^j$, see the top row of the Fig.~\ref{fig:helix_dispersion}. The absence of additional peaks in the spectrum is demonstrated by the dispersion maps below, see bottom raw of Fig.~\ref{fig:helix_dispersion}.

\section{Discussion}
\label{sec:discussion}

We have performed a detailed study of statics and linear dynamics of magnetization in the helix wire with different anisotropy. We have limited our study by hard magnets, which can be well described by the model of anisotropic Heisenberg magnets. Our study was limited by the condition \eqref{eq:hard-validity}.

Let us discuss how our model can be generalized with account of the long range magnetostatics effects. The non-local magnetostatic interaction for thin wires of circular and square cross-sections is known \cite{Slastikov12} to be completely reduced to a local effective easy--tangential anisotropy. It is important that such a conclusion survives for the case of curved wires.\cite{Slastikov12} Thus the magnetostatic interaction can be taken into account as additional anisotropy. In general, one has to consider the model of biaxial magnet. Here we limit ourselves by the helix wire with easy--tangential magnetocrystalline anisotropy. In this case the magnetostatics effects can be taken into account by a simple redefinition of the anisotropy constants, leading to a new magnetic length
\begin{equation} \label{eq:K-eff}
\begin{split}
K&\rightarrow K^{\text{eff}}=K+\pi M_s^2, \\
w&\rightarrow w^{\text{eff}} = \sqrt{\frac{A}{K^{\text{eff}}}} = \frac{2\ell}{\sqrt{1+2Q}}.
\end{split}
\end{equation}
Thus our model \eqref{eq:enegry} is also suitable for thin wires made of a magnetically \emph{soft} material under the restriction
\begin{equation} \label{eq:validity}
h \ll w,\ \ell,\ \frac{1}{\kappa},\ \frac{1}{\tau}.
\end{equation}

In order to check our predictions about effective anisotropy we perform numerical simulations with account of the nonlocal dipolar interaction as described in Sec.~\ref{sec:simulations}. Numerically we integrate Eqs.~\eqref{eq:LL-eq-mod} with the energy \eqref{eq:tot-energy-eq-mod}.

First, we simulate the anisotropic wire with account the dipolar interaction with the energy \eqref{eq:tot-energy-eq-mod}. In this case we need to modify the magnetic length according to \eqref{eq:K-eff}. Thus we also need to redefine the reduced curvature and torsion as follows
\begin{equation} \label{eq:eff-kappa-sigma}
\varkappa\rightarrow \varkappa^{\text{eff}} = \tau w^{\text{eff}}, \quad \sigma\rightarrow \sigma^{\text{eff}} = \tau w^{\text{eff}}.
\end{equation}
For the case $Q=2$, one get $\ell=w$ and $w^{\text{eff}}=2\ell/\sqrt5$. One can see that we have a very nice agreement between the analytical results \eqref{eq:helix-gs-1} and simulations data, see yellow symbols in Fig.~\ref{fig:helix_gs}; we label these data as (am).

The second kind of simulations with account of the dipolar interaction was aimed to verify the validity of our approach for \emph{soft} magnets with $Q=0$. For this purpose we model the soft isotropic wire with account of the dipolar interaction. According to \eqref{eq:K-eff} we get $w^{\text{eff}} = 2\ell$. Simulations data are presented in Fig.~\ref{fig:helix_gs} by dotted symbols [labeled as (m)] for curvature and torsion redefined according to \eqref{eq:eff-kappa-sigma}. By comparing simulations data with analytical results one can see the pretty good agreement in the wide range curvatures and torsions. Our simulations data for soft magnet differ from the theoretical predictions for hard magnets only for relatively high curvature in the vicinity of the boundary with the onion state.

Thus we can conclude that our model of anisotropic Heisenberg magnet is physically sound also for thin wires made of a magnetically soft material.

In conclusion, we have presented a detailed study of statics and linear dynamics of magnetization in the helix wire. We have described  equilibrium magnetization states for three types of uniaxial anisotropy, according to  possible curvilinear directions. All three cases have been realized experimentally in rolled--up ferromagnetic microhelix coils. \cite{Smith11} We have calculated the phase diagram of possible states in case of easy-tangential anisotropy: the quasi-tangential configuration \eqref{eq:helix-gs-1} is energetically preferable for the strong anisotropy case. In this case the deviations from the strictly tangential direction (corkscrew orientation\cite{Smith11}) are caused by the torsion, the direction of the deviation depends on both helix chirality and the magnetochirality of the magnetization structure, see Eq.~\eqref{eq:helix-asymp}. In case of high curvature there is the onion equilibrium state \eqref{eq:onion} in analogues to the onion state in magnetic ring wires \cite{Klaui03a,Guimaraes09}. The magnetization distribution \eqref{eq:helix-gs-binormal} of the quasi-binormal state is directed almost along the binormal (hollow--bar orientation \cite{Smith11}). In contrast to the quasi--tangential state and quasi--binormal one (which are realized for the easy--tangential and easy--binormal magnets, respectively), the normal state for the easy--normal magnets has several peculiarities: (i) it has the form of \emph{exact} normal magnetization distribution along the normal direction $\vec{e}_{\text{\sc{n}}}$, see \eqref{eq:helix-gs-radial}; (ii) the normal state phase is realized for small curvatures and torsions only: $\varkappa^2/\varkappa_0^2 + \sigma^2/\sigma_0^2 < 1$, see Fig.~\ref{fig:helix_phd}(b); (iii) the spectrum of spin waves on the normal state background is symmetric with respect to the direction of the wave propagation.

The torsion of the wire manifests itself in the magnetization dynamics: an effective magnetic field, induced by the torsion breaks the mirror symmetry with the spin wave direction. The dispersion law of spin waves \eqref{eq:disp} is essentially affected by this field.

There is a connection between the helix geometry and the tube one: when the helix pitch vanishes, we have a close-coiled solenoid magnet, which properties are similar to the thin shell nanotube. The spin-wave spectrum in the nanotube is known\cite{Gonzalez10} to have a gap, caused by the curvature. This conclusion is in agreement with the dispersion law for the helix wire, see Fig.~\ref{fig:helix_dispersion}(a). One has to note that the analogy between two systems is adequate under the restriction of vanishing torsions ($\sigma\to0$); this explains the absence of the linear shift in the dispersion law for the nanotube in comparison with \eqref{eq:disp}. In general the transition from 1D systems to 2D requires more accurate account of the dipolar interaction.

We considered the simplest example of the curved wire with constant curvature and torsion. Our results can be generalized for the case of variables parameters $\kappa (s)$ and $\tau (s)$. To summarize we can formulate few general remarks about the curvature and torsion effects in the spin wave dynamics. The linear magnetization dynamics can be described by the generalized Scr{\"o}dinger equation \eqref{eq:Schroedinger}. In case of the straight wire, one has the standard Scr{\"o}dinger equation for the complex magnon amplitude $\psi $ with the typical potential scattering. Th curvature induces an additional effective potential, the `geometrical potential'.\cite{Costa81}  This is described by the modification of effective potential $U$ in Eq.~\eqref{eq:potentials}. Besides, there is a curvature induced coupling potential $W$: the problem becomes different in principle from the usual set of coupled Scr{\"o}dinger equations, see the discussion in Ref.~\cite{Sheka04}. Due to the torsion influence there appears an effective magnetic field. The vector potential of this field is constant for the helix wire, see \eqref{eq:V2-A}, hence the effective magnetic flux density $\vec{B}=\vec{\nabla}\times\vec{A}$ vanishes. Nevertheless the presence of magnetic field with the vector potential $\vec{A}$ breaks the mirror symmetry of the problem: the motion of magnetic excitations in different spatial direction is not identical.

Let us mention the connection between the vector potential and the effective Dzyaloshinskii interaction: the total energy of the Dzyaloshinskii interaction $E_{\mathrm{ex}}^{D}\propto \int \mathrm{d}s \, \vec{A}\cdot\vec{j}$ with the current $\vec{j}= |\psi |^2\vec{\nabla}\arg \psi $, see Eq.~\eqref{eq:A-v-a-ED}. Using an explicit form of the integrand one can find that $E_{\mathrm{ex}}^{D}\propto \sigma q\mathcal{C}$, which reflects the relation between the topology of the wire (namely, helix chirality) with the topology of the magnetic structure (namely, the magnetochirality). In this context it is instructive to note that there is a deep analogy between the Dzyaloshinskii–-Moriya interaction and the Berry phase theory \cite{Freimuth14}.

We expect that our approach can be easily generalized for the arbitrary curved wires, where all potentials becomes spatially dependent: $U(s)$, $W(s)$, and $A(s)$. Depending on the curvature and the torsion these potentials can repel or attract magnons. In latter case there can appear a well with possible bound states, i.e. local modes.

\begin{acknowledgments}
	
The authors thank D. Makarov for stimulating discussions and acknowledge the IFW Dresden, where part of this work was performed, for kind hospitality. The present work was partially supported by the Program of Fundamental Research of the Department of Physics and Astronomy of the National Academy of Sciences of Ukraine (project No. 0112U000056). D.D.S. acknowledges the support from the Alexander von Humboldt Foundation.
	
\end{acknowledgments}

\appendix

\section{Onion-state solution}
\label{app:onion}

We start from the static form of Landau--Lifshitz equations \eqref{eq:LLE-helix}:
\begin{equation} \label{eq:LL-statics}
F\left(\theta,\phi\right)=0, \qquad G\left(\theta,\phi\right)=0
\end{equation}
with $F$ and $G$ being the nonlinear operators,
\begin{equation} \label{eq:LLE-helix-static}
\begin{split}
& F\left(\theta,\phi\right)= - \partial_{\chi\chi} \theta-\sigma \cos \phi  \left( \varkappa  \cos2\theta  - 2 \partial_\chi \phi  \sin^2\theta \right)\\
& +\sin \theta  \cos \theta  \left[\left(\varkappa  + \partial_\chi \phi \right)^2\!\! -(1+\sigma ^2)\! \cos^2\phi  \right],\\
& G\left(\theta,\phi\right)= 
\sin^2\!\theta  \left[-\partial_{\chi \chi}\phi   + (1+\sigma^2)\!\sin \phi \cos \phi \right.\\
& \left.  - 2\sigma \partial_\chi \theta \cos \phi\right] + \sin \theta \cos \theta  \left[\varkappa  \sigma  \sin \phi - 2 \partial_\chi \theta  \left( \varkappa  + \partial_\chi \phi \right) \right].
\end{split}
\end{equation}
By substituting here the expansion \eqref{eq:onion} in the form
\begin{equation} \label{eq:onion-expansion}
\begin{split}
\theta (\chi ) &=  \frac{\pi}{2} + \varepsilon \sum_{n=1}^N \vartheta _n \cos (2n-1)\chi ,\\ 
\phi (\chi ) &= -\chi + \varepsilon \sum_{n=1}^N \varphi _n \sin 2n \chi,
\end{split}
\end{equation}
and expanding results into series over $\varepsilon$ up to the $N$-th order, one get the Fourier expansion of operators $F$ and $G$ as follows
\begin{equation} \label{eq:helix-nonlinear}
\begin{split}
F\left(\theta,\phi\right)&=\sum_{n=1}^N F_n\left(\vartheta_1,\ldots,\vartheta_n;\varphi_1,\ldots,\varphi_n\right)  \cos (2n-1)\chi ,\\
G\left(\theta,\phi\right)&=\sum_{n=1}^N G_n\left(\vartheta_1,\ldots,\vartheta_n;\varphi_1,\ldots,\varphi_n\right)  \sin 2n \chi.
\end{split}
\end{equation}
Here $F_n$ and $G_n$ are polynomials of the order $N$ with respect to $\vartheta_k$ and $\varphi_k$. Then the  Landau--Lifshitz equations \eqref{eq:LL-statics} results in the set of nonlinear polynomial equations
\begin{equation} \label{eq:LL-F-G}
\begin{aligned}
&F_n\left(\vartheta_1,\ldots,\vartheta_n;\varphi_1,\ldots,\varphi_n\right)=0\\
&G_n\left(\vartheta_1,\ldots,\vartheta_n;\varphi_1,\ldots,\varphi_n\right)=0, 
\end{aligned}
\qquad n=\overline{1,N},
\end{equation}
which can be solved numerically on $\vartheta_k$ and $\varphi_k$ with any precision.

In order to calculate the energy of the onion state, we substitute the magnetization angles $\theta$ and $\phi$ in the form \eqref{eq:onion-expansion} into the energy density \eqref{eq:energy-angular}, expand the results over $\varepsilon$ up to the $2N$-th order and average the result over the helix period,
\begin{equation} \label{eq:En-onion}
\begin{split}
&\mathscr{E}^{\mathrm{on}}(\sigma,\varkappa)= \frac{1}{2\pi}\int_0^{2\pi} \mathscr{E}\mathrm{d}\chi,\\
&\mathscr{E} = \mathscr{E}_{\text{ex}} + \mathscr{E}_{\text{an}}^{\mathrm{ET}} = \mathscr{E}\left(\vartheta_1,\ldots,\vartheta_n;\varphi_1,\ldots,\varphi_n\right).
\end{split}
\end{equation}

\bigskip

%
%----------------------------------------------------------------
%
%\bibliography{soliton}

%
%\include{reply}

\end{document}